# Coronal Heating and Solar Wind Generation by Flux Cancellation Reconnection

D. I. Pontin[1] , E. R. Priest[2] , L. P. Chitta[3] , and V. S. Titov[4]
[1] School of Information and Physical Sciences, University of Newcastle, Callaghan, NSW 2308, Australia; david.pontin@newcastle.edu.au
[2] School of Mathematics and Statistics, University of St Andrews, St Andrews, KY16 9SS, UK
[3] Max Planck Institute for Solar System Research, Justus-von-Liebig-Weg 3, D-37077, Göttingen, Germany
[4] Predictive Science Inc., 990 Mesa Rim Road, Suite 170, San Diego, CA 92121, USA

## Abstract

In this paper, we propose that flux cancellation on small granular scales (≲1000 km) ubiquitously drives reconnection at a multitude of sites in the low solar atmosphere, contributing to chromospheric/coronal heating and the generation of the solar wind. We analyze the energy conversion in these small-scale flux cancellation events using both analytical models and three-dimensional, resistive magnetohydrodynamic (MHD) simulations. The analytical models—in combination with the latest estimates of flux cancellation rates—allow us to estimate the energy release rates due to cancellation events, which are found to be on the order $10^6$–$10^7$ erg cm$^{-2}$ s$^{-1}$, sufficient to heat the chromosphere and corona of the quiet Sun and active regions, and to power the solar wind. The MHD simulations confirm the conversion of energy in reconnecting current sheets, in a geometry representing a small-scale bipole being advected toward an intergranular lane. A ribbon-like jet of heated plasma that is accelerated upward could also escape the Sun as the solar wind in an open-field configuration. We conclude that through two phases of atmospheric energy release—precancellation and cancellation—the cancellation of photospheric magnetic flux fragments and the associated magnetic reconnection may provide a substantial energy and mass flux contribution to coronal heating and solar wind generation.

*Unified Astronomy Thesaurus concepts:* Solar coronal heating (1989); Solar chromospheric heating (1987); Solar magnetic reconnection (1504); Solar wind (1534); Solar physics (1476); Solar magnetic fields (1503)

## 1. Introduction

Recent observations from the Solar Orbiter and Parker Solar Probe missions have shed new light on the important role of small-scale magnetic reconnection in coronal heating and in the generation of the solar wind. Reconnection is thought to produce two newly identified phenomena. The first are small-scale extreme ultraviolet (EUV) brightenings termed "campfires" in the low corona at heights of 1000–5000 km above the solar surface (Berghmans et al. 2021; Zhukov et al. 2021). The second, observed far from the solar surface out in the solar wind, are ubiquitous apparent reversals of the radial magnetic field, termed "switchbacks," with potential origins from supergranular- to granular-scale reconnection dynamics in the low corona (Bale et al. 2021, 2023). Reconnection between closed–closed or closed–open magnetic fields are commonly invoked drivers of both the campfires and switchbacks (Chen et al. 2021; Tripathi et al. 2021; Raouafi et al. 2023). Here, we propose a unified model (Figure 1) for coronal heating and solar wind generation that builds on these recent studies. We suggest that flux cancellation on small granular scales of ≲1000 km (Smitha et al. 2017) ubiquitously drives reconnection at a multitude of sites. This work extends previous studies by analyzing flux cancellation reconnection in detail, including the release of energy and its dependence on the fluxes involved.

Importantly, photospheric flux cancellation is now known to be very much more common than thought previously. Observations from the SUNRISE balloon mission (Solanki et al. 2017) revealed the photospheric magnetic field at a spatial resolution of 0″.15 and revealed magnetic flux emerging and canceling on granular scales. Smitha et al. (2017) tracked magnetic features in the quiet Sun with fluxes of $10^{15}$–$10^{18}$ Mx and discovered a flux emergence and cancellation rate an order of magnitude higher than previous estimates, namely, 1100 Mx cm$^{-2}$ day$^{-1}$.

Reconnection driven by magnetic flux cancellation has long been associated with a wide range of phenomena, such as X-ray bright points (Martin et al. 1985; Priest et al. 1994; Archontis & Hansteen 2014) and X-ray jets (Shibata et al. 1992; Shimojo et al. 2007) in the corona; Hα Ellerman bombs in the low chromosphere near sunspots or in the quiet Sun (Rouppe van der Voort et al. 2016; Hansteen et al. 2017); UV bursts in the active-region chromosphere (Peter & Dwivedi 2014); and explosive events in the transition region (Brueckner & Bartoe 1983; Innes et al. 1997). However, it is only with the realization from SUNRISE that flux cancellation is much more common that it has been thought to be a viable mechanism for heating the whole corona (Priest et al. 2018), and this is supported by a range of other observations. There are clear examples of flux cancellation triggering brightening in coronal loops (Tiwari et al. 2014; Huang et al. 2018), some in an active region associated also with a UV burst and bidirectional jets (Chitta et al. 2017b). Furthermore, flux cancellation in regions of complex mixed polarity has been observed in association with brightenings in the cores of active regions (Chitta et al. 2018, 2020). Also, evidence has been presented that most of the small-scale campfires are formed by magnetic reconnection driven by flux cancellation (Panesar et al. 2021), which occurs either between two main footpoints of a bipolar feature or between one of those footpoints and a nearby magnetic fragment of opposite polarity (Kahil et al. 2022).

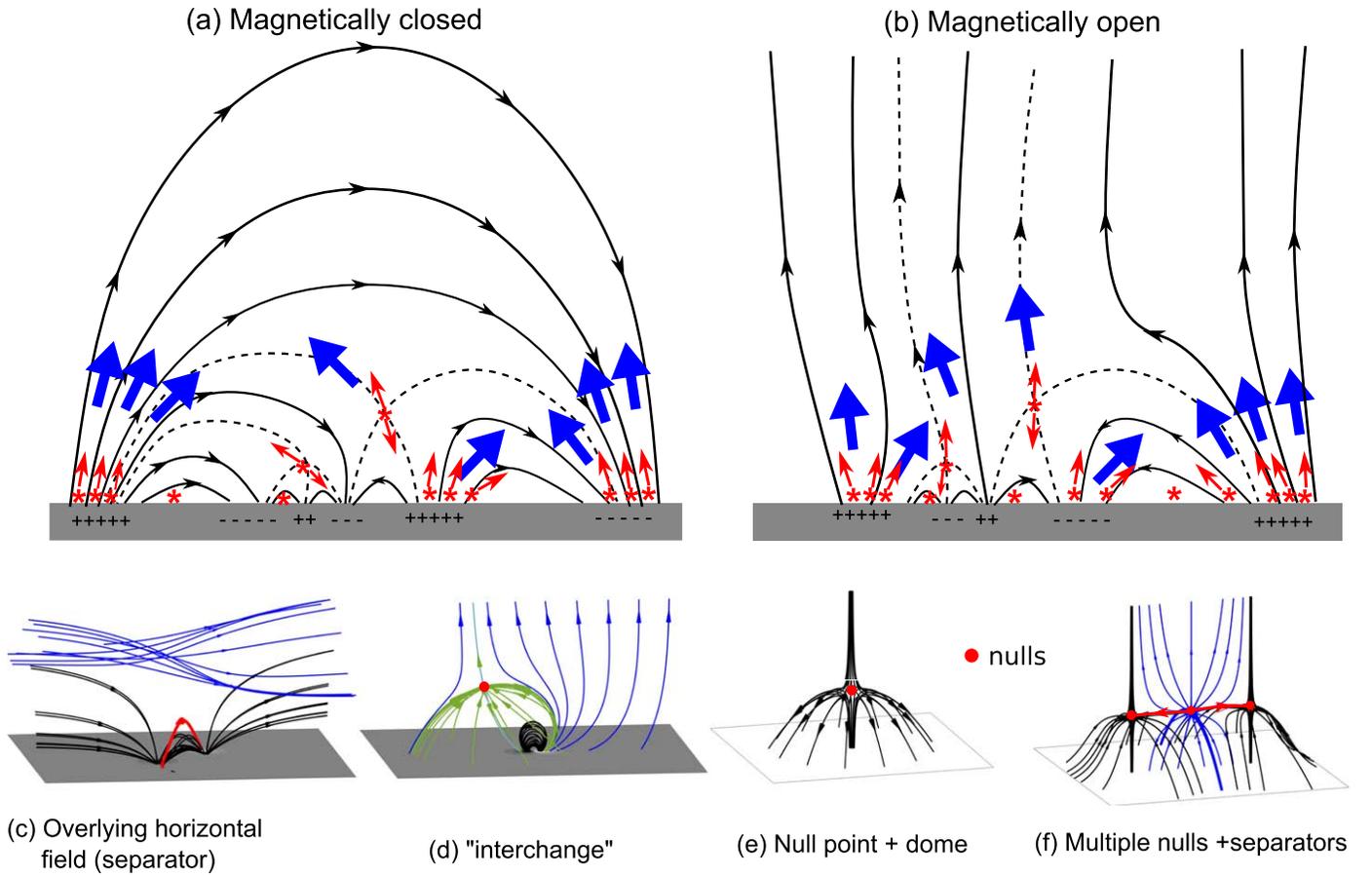

**Figure 1.** A unified model for coronal heating and solar wind generation in regions above supergranules that are (a) magnetically closed and (b) magnetically open. Stars indicate reconnection sites in the lower atmosphere that generate flows (red arrows) as well as fast particles, waves, and heat flux (blue arrows). Panels (c), (d), (e), and (f) indicate close-ups of different types of reconnection taking place locally at much smaller granular scales, at nulls or separators (red) in the atmosphere. Red dots indicate null points, thick red lines are separator field lines, and the green field lines in (d) show the separatrix dome of a magnetic null.

As far as the solar wind is concerned, three new sets of observations are important, namely those of transition-region upflows, those of *interchange reconnection/jetlets*, and possibly those of *switchbacks*. While the first and second of these are relevant for the generation of the solar wind, the third one describes the features within the solar wind itself. Considering the former first, Tripathi et al. (2021) analyzed transition-region intensity, Doppler, and nonthermal velocities in the quiet Sun and coronal holes, as observed by IRIS as a function of magnetic flux density. Based on these observations, they proposed a simple scenario for heating the corona and driving the solar wind, by interchange reconnection in coronal holes and by reconnection between closed-field lines in the quiet Sun. The observations were extended to include chromospheric lines by Upendran & Tripathi (2022), who found that coronal holes have reduced intensity in all of the lines; the chromospheric lines have excess flows in coronal holes, while the transition region shows excess upflows in coronal holes and downflows in the quiet Sun. The role of interchange reconnection in the formation of the solar wind has also recently been supported by simulations. Iijima et al. (2023) simulate from the convection zone out to 29 $R_\odot$ with a solar wind driven by turbulent convection. They find that around half of the upward magnetic energy flux in the open field region is supplied by cross-field transport from the closed field.

We turn now to the observations of transition region network jets and coronal jetlets. When flux cancellation occurs within the open field regions of coronal holes (or on their peripheries) it drives interchange reconnection, i.e., reconnection between open and closed magnetic flux (Fisk et al. 1999). Such interchange reconnection could occur on large active-region scales (Baker et al. 2023; Chitta et al. 2023a), or for our purposes here, on very small granular scales. It produces plasma jets, energetic transient phenomena observed across a wide range of scales (Tian et al. 2014; Raouafi et al. 2016). The reconnection, and the jets, may be triggered by a number of mechanisms including flux cancellation but also flux emergence (e.g., Yokoyama & Shibata 1996; Moreno-Insertis & Galsgaard 2013), kink-type instabilities (e.g., Pariat et al. 2009; Wyper et al. 2016), and the eruption of low-lying filaments in the closed field (e.g., Antiochos et al. 1999; Sterling et al. 2015; Wyper et al. 2017). Based on the statistics of polar X-ray jets detected by Hinode, it was previously thought that they do not contribute substantially to the solar wind mass and energy flux (e.g., Paraschiv et al. 2015; Lionello et al. 2016). However, it appears that smaller and cooler jets may be much more numerous, and recently there has been renewed attention on these small-scale jets ("jetlets"), which are suggested to be ubiquitous in both open- and closed-field regions. Wang et al. (2022) identified 88 flux cancellation events associated with H$\alpha$ spicules, out of which seven events are associated with EUV jets/spicules. Based on the statistics of the observed jets (that are on scales of a few hundred kilometers), Raouafi et al. (2023) argued that they can provide sufficient mass and energy flux into the corona to explain the solar wind (see also Chitta et al. 2023b). We therefore propose that such jets are due to small-scale flux cancellation reconnection driven by photospheric flux cancellation on granular scales.

Finally, one of the most striking new results from Parker Solar Probe and Solar Orbiter is the prevalence of *switchbacks*: localized regions where the magnetic field vector rotates such that the radial magnetic field is greatly depleted or even reversed (Bale et al. 2019; Kasper et al. 2019). The relevance for the present work is that one leading candidate mechanism for the formation of switchbacks is interchange reconnection in the low corona. The reconnection may either locally form kinked field lines that are carried outward by the solar wind (Fisk & Kasper 2020; Zank et al. 2020), or the switchbacks may be remnants of flux ropes ejected outward from the reconnecting current layer (Drake et al. 2021). On the other hand, recent simulations of interchange reconnection in three dimensions by Wyper et al. (2022) suggest that the field rotations produced in the high corona are modest, because flux ropes in 3D have a tendency to untwist. Nevertheless, they show that bursty interchange reconnection launches Alfvénic fluctuations into the open field, which may form the seeds for the growth of switchbacks, for example by turbulent steepening (see, e.g., Squire et al. 2020), and may help to explain the "patchy" nature of the switchback observations.

The outline of the paper is as follows. First, in Section 2, we introduce the underlying physics of our model. In Section 3, we present some observational context for quiet-Sun coronal jets. In Section 4, we propose an analytical model. In Section 5, we estimate the energy release. In Sections 6.1 and 6.2, we describe the setup and parameters for our 3D magnetohydrodynamic (MHD) simulation, and then the results. We finish in Section 7 with discussion and conclusions.

## 2. Flux Cancellation Reconnection Model

We propose a model in which photospheric flux cancellation on granular scales drives magnetic reconnection in the low corona, with the associated released energy heating the plasma and driving the solar wind. Flux cancellation reconnection converts stored magnetic energy to kinetic, thermal, and other forms of energy. In steady MHD models, typically three-fifths of the energy is released as kinetic energy and two-fifths as heat (Priest 2014), but some energy may also be converted to accelerate particles and some to MHD waves when the reconnection is nonsteady (Longcope & Priest 2007). The heat, fast particles, and waves that are generated are transported away from a reconnection site. In a magnetically closed region above a supergranule (Figure 1(a)), the kinetic, fast-particle, and wave energy is subsequently degraded and converted into heat by viscous or shock effects, so that most of the released energy goes into heating the plasma. On the other hand, in a magnetically open region (Figure 1(b)), the flows that are generated are free to escape upward along the open magnetic field lines; they could show up as network jets or jetlets that could act as sources for the solar wind. The result is that the magnetically open regions are heated to lower temperatures than closed regions, and their plasma densities are also lower because plasma escapes upward. The effect is also enhanced if the open regions have smaller field strengths which reduces the reconnection-produced heating.

Magnetic reconnection is taking place on granular scales in different ways, depending on the local topology, as indicated in Figures 1(c)–(f). If the local field has a dominant horizontal component and two opposite-polarity fragments are approaching one another, they will drive separator reconnection (Figure 1(c)). If, on the other hand, a concentrated isolated polarity moves within a dominant vertical field of the opposite polarity, then null-point reconnection will be driven in a separatrix dome (Figure 1(e)). If the isolated polarity is more elongated, then three or more null points can occur in the overlying dome, joined by separators, and now separator reconnection can be driven (Figure 1(f)). The reconnection could occur at the footpoint of a coronal loop or within a coronal hole. If magnetic flux emerges as a local (magnetically connected) bipole, and interacts with opposite-polarity field at, for instance, the boundary of a supergranule cell—Figure 1(d)—it can drive interchange reconnection, which in practice occurs again at the nulls or separators (if they are present) of an overlying separatrix dome structure (green in Figure 1(d)).

We suggest that flux cancellation reconnection occurs in two phases:

(i) A *Precancellation Phase*, during which opposite-polarity photospheric magnetic fragments approach one another and drive reconnection in the overlying atmosphere when the fragments are closer than the *flux interaction distance* $[F_0/(\pi B_0)]^{1/2}$ (Longcope 1998), where $F_0$ is the magnitude of the flux fragment and $B_0$ is the strength of the overlying magnetic field. The height of the reconnecting current sheet depends on the values of $F_0$ and $B_0$, as well as the distance that the fragments have moved (from some initial state in which the field is assumed to be potential; Priest et al. 2018).

(ii) A *Cancellation Phase*, in which the flux fragments actually cancel in the photosphere by reconnection (Priest & Syntelis 2021). It is not yet clear how much of the energy that is released in this phase, if any, can escape to the overlying atmosphere, but Ellerman bombs near sunspots or in the quiet Sun (Rouppe van der Voort et al. 2016; Hansteen et al. 2017) are likely to occur during this phase when the reconnection is occurring very low in the atmosphere.

In Sections 4–6, we complement our previous work on reconnection in horizontal fields (e.g., Priest et al. 2018) by describing models for reconnection with a single coronal null in an open magnetic field such as the "interchange" configuration, corresponding to Figures 1(d) and (e). In what follows, we model principally the precancellation phase, for simplicity. Simulations of the cancellation phase itself will be described in a follow-up work.

At this point, it is worth stressing again that interchange reconnection and jetlets can additionally be triggered by processes other than flux cancellation, such as flux emergence. Indeed, Wang (2020) has used Solar Dynamics Observatory observations of emerging ephemeral regions to derive energy fluxes consistent with the solar wind, and the same process may additionally be important for heating closed-field regions (Wang 2022). Also, we note that interchange reconnection—however it is triggered—may transport energy from closed to open field regions in the form of laminar or turbulent fluctuations, which may contribute to the acceleration of the outgoing wind (Zank et al. 2018; Adhikari et al. 2020).

## 3. Observations of Small-scale Coronal Jets

To provide observational context to the model presented here, we show examples of quiet-Sun coronal jets in the EUV in Figure 2. These data were obtained by the 174 Å EUV High

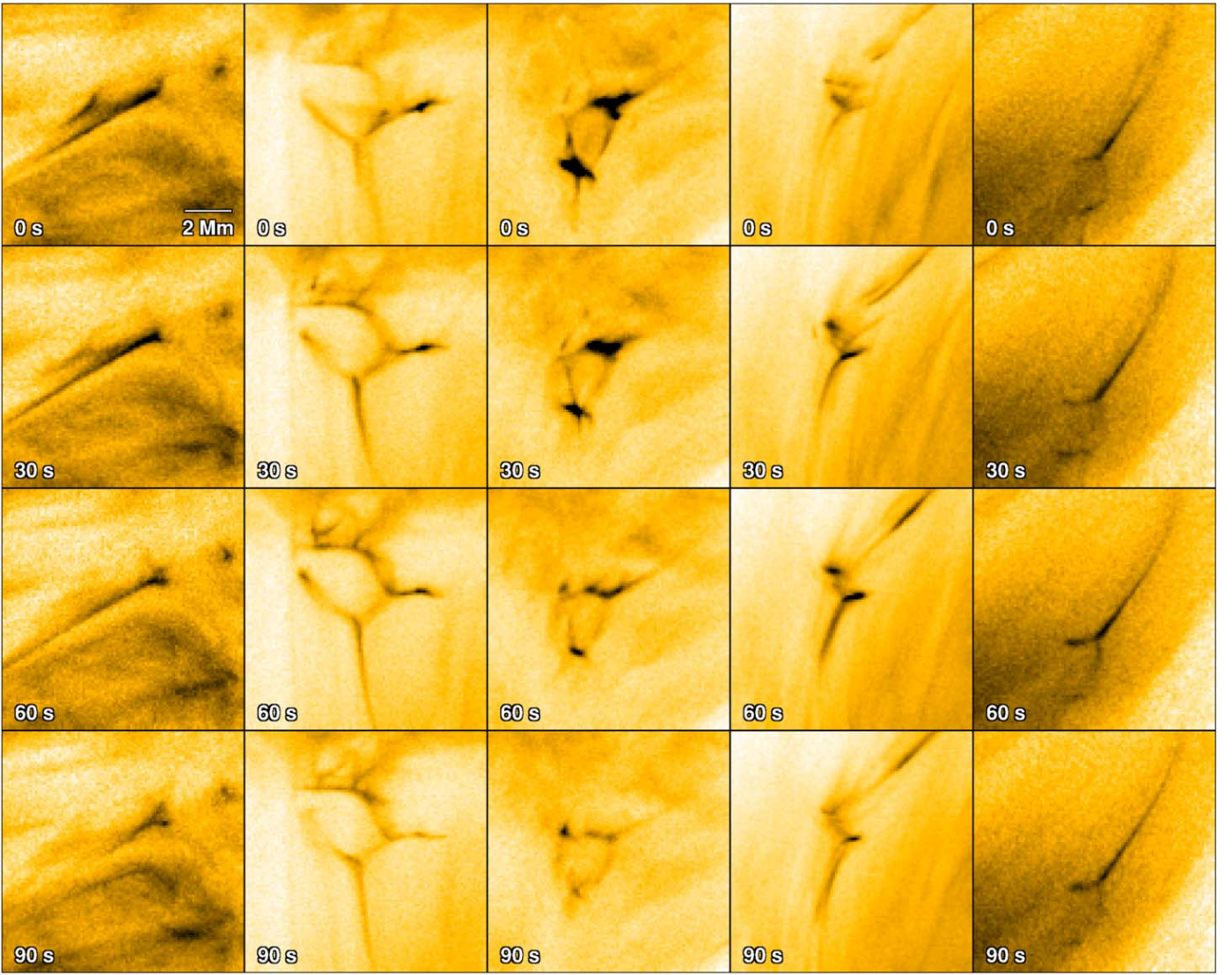

**Figure 2.** Five small-scale jets in the quiet solar corona. Each column shows snapshots at 30 s intervals from a time sequence of observations of five Y-shaped coronal jets recorded by the HRI$_{EUV}$ instrument on Solar Orbiter (the inverse intensity is plotted). Time runs from top to bottom. Each panel covers a field of view of 10.55 Mm × 10.55 Mm, and for reference, a 2 Mm scale is overlaid on the top left panel. The images are plotted on a negative color scale.

Resolution Imager (HRI$_{EUV}$) of the Extreme Ultraviolet Imager (EUI; Rochus et al. 2020) on Solar Orbiter (Müller et al. 2020). The observations were recorded on 2022 October 12 during one of the second science perihelion campaigns of Solar Orbiter. At the time of observations, Solar Orbiter was at a distance of 0.293 au from the Sun. At these distances, HRI$_{EUV}$ has a spatial resolution of about 210 km. Thus, the presented data are among the highest spatial resolution EUV observations ever made of the quiet Solar corona. The jets show Y-shaped morphology with a narrow elongated spire and a broader, brighter base. Thanks to these high-resolution data, we are now able to directly observe coronal jets with base widths down to 1 Mm (i.e., the typical size of a solar granule; see the left panel of Figure 2). These examples point to the role of small-scale magnetic processes in coronal heating. In a recent study, Chitta et al. (2023b) found ubiquitous reconnection-driven Y-shaped jet activity in a coronal hole with kinetic energy fluxes as low as $10^{21}$ erg, and suggested that they could contribute substantially to the solar wind mass and energy flux, which reinforces our proposal above regarding the cause of the solar wind.

## 4. Analytical Model

In this section, we present an analytical model that is a 2D representation of the fully 3D configuration, for tractability (we extend this to full 3D through simulation in Section 6). Consider a main source of negative flux at the origin situated within a uniform vertical field of strength $B_0$ (Figure 3(a)). The initial 2D magnetic field is

$$\boldsymbol{B} = B_x\hat{\boldsymbol{x}} + B_y\hat{\boldsymbol{y}} = -\frac{F_0}{\pi}\frac{\hat{\boldsymbol{r}}}{r} + B_0\hat{\boldsymbol{y}}$$
$$= B_0\left(-\frac{h_0 x\hat{\boldsymbol{x}} + h_0 y\hat{\boldsymbol{y}}}{x^2 + y^2} + \hat{\boldsymbol{y}}\right), \quad (1)$$

where $r$ is the distance from the origin to the point $(x, y)$, $F_0$ is the flux of a 2D source, and $h_0 = F_0/(\pi B_0)$ is the initial height of the null, which is obtained by setting $x = 0$ and $B_y = 0$ in Equation (1) ($B_x$ being zero at $x = 0$ by symmetry).

When the main source of negative flux moves, it drives reconnection (spine-fan reconnection in full 3D;

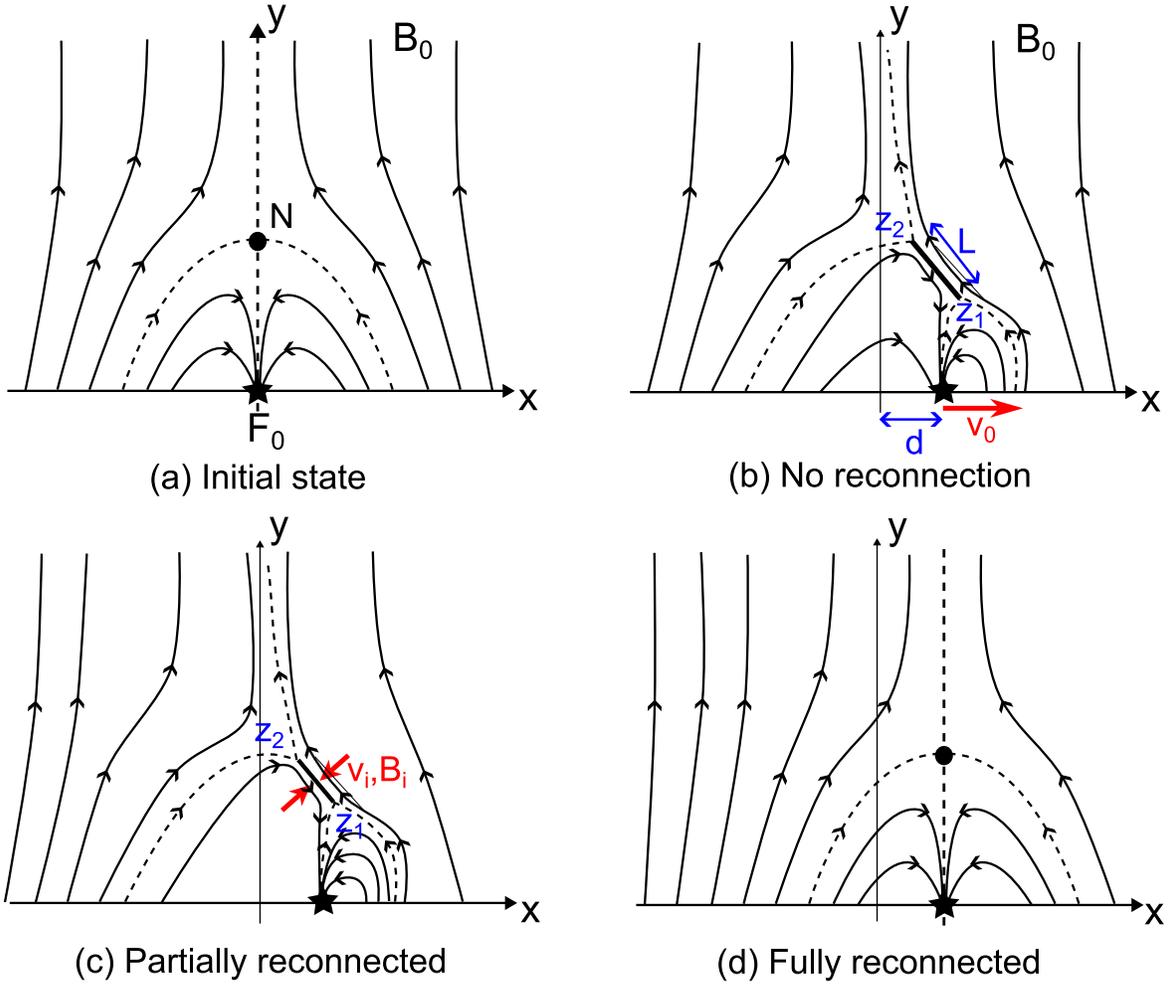

**Figure 3.** (a) The initial magnetic field for an isolated source of negative flux ($F_0$) situated in a uniform vertical field of strength $B_0$. The field has a null point, $N$, at the summit of a separatrix dome (dashed) enclosing the negative flux. (b) The magnetic field after the negative source has moved a distance $d$ to the right at speed ($v_0$). If there is no reconnection, a current sheet of length $L$ is created. (c) When the field has partially reconnected, a reconnecting current sheet forms with inflowing magnetic field $B_i$ at speed $v_i$. (d) When the field has completely reconnected, the configuration is the same as the initial one, but translated along a distance $d$.

Pontin et al. 2007, 2013) at a current sheet that forms about the null point (Figure 3). If there is no reconnection, a long current sheet is created, as shown in Figure 3(b), and the positions $z_1$ and $z_2$ of the ends of the sheet can be calculated by using flux conservation. Initially (Figure 3(a)), the two fluxes below the dome are each equal to $F_0/2$, and the vertical flux to the right or left of the null point is, say, $F_\infty$. When the main source denoted by the star in Figure 3 has moved a distance $d$ to the right, we assume that flux is conserved at the source and at infinity, and across the line $C2$ shown in Figure 4, while the fluxes in the dome below $z_1$ and $z_2$ have decreased and increased, respectively, by $B_0 d$ per unit distance out of the plane, for reasons described in Appendix D. The flux conditions along $C1$ and $C2$ (see Figure 4) are then, respectively,

$$F_{C1} \equiv \int_{y=0}^{y=y_1} B_x(x, y) dy = \tfrac{1}{2} F_0 - B_0 d, \quad (2)$$

$$F_{C2} \equiv \int_{x=x_1}^{x=\infty} B_y(x, y_1) dx = F_\infty, \quad (3)$$

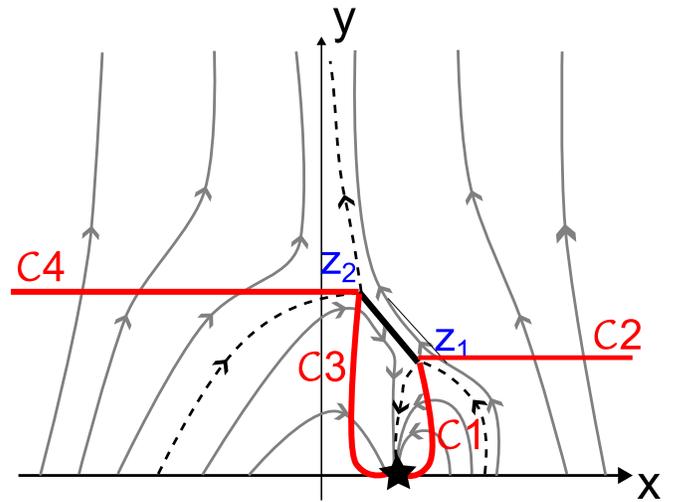

**Figure 4.** The lines $C1$, $C2$, $C3$, and $C4$ across which the fluxes are calculated. $C1$ extends down to the $x$-axis from the endpoint $z_1$, provided $z_1$ is to the right of the source, and $C3$ extends down from endpoint $z_2$ to the $x$-axis, provided $z_2$ is to the left of the source. $C2$ extends from $z_1$ to the right to infinity, and $C4$ extends to the left from $z_2$.

where the expression for $F_{C1}$ assumes $x_1 > d$ and is calculated along a curved line from the main source to $z_1$.

Now suppose instead that the flux has partially reconnected, by being carried into the sheet from both sides at speed $v_i$ with strength $B_i$, as shown in Figure 3(c), in such a way that the amount of flux that has been reconnected while the main source has moved a distance $d$ in time $\tau$ at speed $v_0 = dd/dt$ is

$$F_R \equiv \int_{t=0}^{t=\tau} v_i B_i dt = \int_{d=0}^{d} (v_i B_i / v_0) dd \qquad (4)$$

per unit distance out of the plane. The reconnection transfers this magnetic flux ($F_R$) from the "inflow" flux domains (lower left and upper right) to the "outflow" flux domains (lower right and upper left), so that comparing Figures 3(b) and (c) it can be seen that there are fewer field lines in the inflow domains and more in the outflow domains after reconnection. These flux domains are partitioned by the dashed lines in Figure 4, such that the inflow domains contain the curves $C2$ and $C3$, while the outflow domains contain $C1$ and $C4$, across which the fluxes may be evaluated. Therefore, the two flux conditions that help determine the endpoints of the current sheet are modified from the expressions in Equation (3) to become

$$F_{C1} = \tfrac{1}{2} F_0 - B_0 d + F_R, \qquad (5)$$

$$F_{C2} = F_\infty - F_R. \qquad (6)$$

If the flux has completely reconnected, then we simply recover the original configuration translated along a distance $d$, as shown in Figure 3(d), in which case the fluxes in each region below the dome are simply $\tfrac{1}{2} F_0$.

The easiest way to model the magnetic field with a current sheet in 2D is to use complex variable theory, and so to write the components of the initial field (1) in the form

$$\mathcal{B}_0(z) \equiv B_y + i B_x = B_0 \left(1 - \frac{ih_0}{z}\right) = B_0 \left(1 - \frac{z_0}{z}\right), \qquad (7)$$

where $z = x + iy$ is the complex number and $z_0 = ih_0$ is the position of the null point. When the source has moved a distance $d$ to $z = d$, we need a magnetic field containing a current sheet stretching from $z = z_1$ to $z = z_2$ that behaves like $-ih_0 B_0/(z-d)$ near $z = d$, and has the correct behavior at infinity. For this purpose, we follow Titov (1992) and use

$$\mathcal{B} \equiv B_y + i B_x = P(z) Q(z), \qquad (8)$$

where $Q = [(z-z_1)(z-z_2)(z-\bar{z}_1)(z-\bar{z}_2)]^{1/2}$ gives two current sheets, one from $z_1$ to $z_2$ and an image one below the photosphere from $\bar{z}_1$ to $\bar{z}_2$, with $z_1 = x_1 + iy_1$ and $\bar{z}_1 = x_1 - iy_1$. The purpose of the image sheet is to preserve the normal field component at the photosphere ($y = 0$). Also,

$$P(z) = -\frac{i}{\pi} \int_{-\infty}^{+\infty} \frac{B_y(\xi, 0)}{(\xi - z) Q(\xi)} d\xi \qquad (9)$$

is an analytic function chosen to satisfy the boundary condition on the $x$-axis.

The values of $z_1$ and $z_2$, and so of $L$, are determined by asymptotic and flux conditions, as detailed in Appendix E and summarized as follows. We first substitute for $B_y(\xi, 0)$ and assume the sheet is short by expanding $z_1$ and $z_2$ in powers of $d \ll h_0$, as

$$z_1 = \tilde{z}_0 + 2\sqrt{dh_0}\, u,$$
$$z_2 = \tilde{z}_0 - 2\sqrt{dh_0}\, u, \qquad (10)$$

where $\tilde{z}_0 = ih_0 + dv$ is the center of the current sheet. Then the condition that the field behaves correctly at infinity (namely, $\mathcal{B} \approx B_0(1 - ih_0/z)$) leads to the conclusion that

$$v = 1 - w^2, \qquad u = (1-i) w / \sqrt{2}, \qquad (11)$$

in terms of a real unknown $w$.

Next, the magnetic field is rewritten as

$$\mathcal{B} = -\frac{iB_0}{h_0} (z - z_1)^{1/2} (z - z_2)^{1/2} f(z)$$
$$= -\frac{B_0 \sqrt{[z - \tilde{z}_0]^2 + 4iw^2 d}}{ih_0 + d}$$
$$\times \left( \frac{ih_0 + (1 - w^2) d}{z - d} + \frac{w^2 d}{z + ih_0} \right), \qquad (12)$$

where $f(z)$ has been expanded in powers of $d$. This may be used to calculate a complex flux function that automatically includes the current sheet between $z_1$ and $z_2$, namely

$$\mathcal{A}(z) = \int \mathcal{B}\, dz. \qquad (13)$$

The final condition is to evaluate this at $\tilde{z}_0$ and equate the value of the real flux function to its value initially at the null point ($z = z_0$), which ensures that there has been no reconnection, because the field line through the initial null point still passes through the end of the current sheet. This determines the value of $w$ as $w = 1$, so that the current sheet is centered at $ih_0$, is inclined at $\pi/4$, and has a length of $L = 4\sqrt{dh_0}$.

Next, we calculate the values of the inflow magnetic field ($B_i$) and inflow speed ($v_i$) at the current sheet and use them to estimate the rate of energy release. The inflow field ($B_i$) is simply the value of the magnetic field at $z = \tilde{z}_0$ in Equation (12), namely

$$B_i = \frac{B_0 L}{2 h_0}. \qquad (14)$$

When $L$ is no longer small, the asymptotic and flux conditions may in principle be used to determine $z_1$ and $z_2$, and therefore $L$.

Now, in order to calculate the inflow speed ($v_i$), let us consider the magnetic flux $\psi$ in the part of the dome below $z_1$, namely $\psi = F_{C1} \bar{L} = \left( \tfrac{1}{2} F_0 - B_0 d + F_R \right) \bar{L}$, where $\bar{L}$ is the extension out of the plane of the current sheet. Thus, because $F_0$ is constant and $F_R = \int v_i B_i dt$, the rate of change of flux due to reconnection is just $d\psi/dt = v_i B_i \bar{L}$. If we now assume the current sheet is small, so that the configuration is close to the final state shown in Figure 3(d), then the change in flux is just $B_0 d\bar{L}$, and its rate of change is simply $B_0 v_0 \bar{L}$. Equating these two expressions for $d\psi/dt$ gives

$$v_i B_i \bar{L} = B_0 v_0 \bar{L} \quad \text{or} \quad \frac{v_i}{v_0} = \frac{B_0}{B_i} = \frac{2 h_0}{L}. \qquad (15)$$

At this point, it is worth emphasizing that the subscript $i$ in the above equation denotes values at the inflow to the current sheet, while the subscript 0 denotes parameters imposed at $t = 0$ (not to be confused with values in the current sheet outflow): specifically, $v_0$ is the speed of the parasitic polarity, $B_0$ is the strength of the vertical field, and $h_0$ is the initial height of the null.

Finally, the value of $L$ may be obtained by using the condition (Priest et al. 2018; Syntelis et al. 2019) that the inflow speed ($v_i$) equal a fraction $\alpha$ of the inflow Alfvén speed ($v_{Ai}$), so that

$$v_i = \alpha \frac{B_i}{B_0} v_{A0}. \quad (16)$$

After substituting for $v_i$ and $B_i$ from Equations (15) and (14), this gives

$$\frac{L^2}{h_0^2} = \frac{4 M_{A0}}{\alpha}, \quad (17)$$

where $M_{A0} = v_0/v_{A0}$ is the Alfvén Mach number of the driving flow.

The rate of energy release at the current sheet is then (Priest et al. 2018)

$$\frac{dW}{dt} = 0.8 \frac{v_i B_i^2}{\mu} L \bar{L}, \quad (18)$$

or after substituting the expressions for $v_i$, $B_i$, and $L$ from Equations (15), (14), and (28),

$$\frac{dW}{dt} = 0.8 \frac{v_0 B_0^2}{\mu} \frac{2 M_{A0}}{\alpha} h_0 \bar{L}. \quad (19)$$

Roughly three-fifths of this energy is expected to be released as kinetic energy and two-fifths as heat, based on steady 2D reconnection models, and this turns out to be consistent with our simulation results (see Section 6).

## 5. Estimates of Energy Release

Three methods may be used to estimate the energy release, all of which support the idea that flux-cancellation-driven magnetic reconnection is a viable mechanism for heating the corona and helping to generate the solar wind.

### 5.1. Phase 1 from the Model

First of all, on the basis of the above analytical model for flux cancellation with an overlying vertical field, the energy input from phase 1 of flux cancellation according to Equation (19), namely the precancellation phase when flux fragments are approaching one another, can be estimated as follows. Here we use the 2D analysis to give an approximation to the 3D magnetic field in the current sheet inflow, noting that the 3D nature of the current sheet is more complex to model but a method to do so has been developed by Priest & Syntelis (2021). We denote the flux of a source in 3D by $F_0^{3D}$ and the height of the null by $h_0^{3D}$, and assume $\bar{L} \approx \pi h_0^{3D}$ (because the current sheet is roughly semicircular and of height $h_0^{3D}$). Then, after putting $h_0^{3D} = \sqrt{F_0^{3D}/(\pi B_0)}$ for the three-dimensional flux interaction distance (Longcope 1998; Priest et al. 2018), the rate of energy release can be written as

$$\frac{dW}{dt} = 1.6\pi \frac{v_0 B_0}{4\pi^2} \frac{M_{A0}}{\alpha} F_0^{3D}, \quad (20)$$

or, if $\alpha \approx 0.1$, $v_4$ is the value of the photospheric driving speed $v_0$ of the flux fragments in units of $10^4$ cm s$^{-1}$, $B_1$ is the ambient vertical magnetic field in units of 10 Gauss, and $F_{18}$ is the flux of the canceling flux patches in units of $10^{18}$ Mx,

$$\frac{dW}{dt} = 4 \times 10^{22} \pi v_4 \, B_1 \, F_{18} \, M_{A0} \text{ erg s}^{-1}. \quad (21)$$

For typical values in the quiet Sun, we assume $v_0 = 1$ km s$^{-1}$, $F_0^{3D} = 3 \times 10^{17}$ Mx, $B_0 = 10$ G, and $M_{A0} = 0.1$, so that an area of $A = 10^{16}$ cm$^2$ is swept out in a time of, say, $10^3$ s, giving a heating per unit area of $4 \times 10^6$ erg cm$^{-2}$ s$^{-1}$, which is sufficient to heat the chromosphere. (If instead $F_0^{3D} = 2 \times 10^{16}$ Mx and $B_0 = 100$ G in the quiet Sun (see below), then we obtain the same heating.) On the other hand, for active regions, if the vertical field is $B_0 = 30$ G and the flux is $F_0^{3D} = 10^{18}$ Mx, the heating becomes $4 \times 10^7$ erg cm$^{-2}$ s$^{-1}$, which is again high enough to heat the chromosphere. Also, if 10%–20% of these values leak to higher levels, they are also sufficient to heat the corona.

### 5.2. Phase 2 from Observed Flux Cancellation

Second, the energy released during phase 2 when the flux is canceling may be estimated as follows. Consider a situation where two flux patches of flux $F$, dimension $D$, and field strength $B$ are canceling, such that the dimension is

$$D = \left(\frac{F}{B}\right)^{1/2} 10^{-5} \text{ km}, \quad (22)$$

where $F$ is measured in Mx and $B$ in Gauss. If the flux patches are squares, then $D$ is just the side of a square. However, if they are circles, then $D$ represents their diameter and the formula is true in order of magnitude, but it becomes exact if the right-hand side is multiplied by a factor of $\sqrt{4/\pi} \approx 1.1$. Thus, for example, if the flux were $10^{15}$, $10^{16}$ or $10^{18}$ Mx, the dimension would be $D = 30$, 100, or 1000 km, respectively, for a field strength of 100 G. Also, the time $\tau$ to cancel the flux at a speed $v$ km s$^{-1}$ would be

$$\tau = \frac{D}{v} \text{ s}. \quad (23)$$

The observed flux cancellation rate in the quiet Sun is 1100 Mx cm$^{-2}$ day$^{-1}$ (Smitha et al. 2017), which amounts to $1.3 \times 10^{14}$ Mx Mm$^{-2}$ s$^{-1}$. This implies that, in one granule occupying an area of roughly 1 Mm$^2$, the typical flux cancellation over the approximately 300 s lifetime of the granule is $4 \times 10^{16}$ Mx. So, let us estimate how much magnetic energy is likely to be converted during this time, depending on the sizes of the flux patches.

In practice, there will be a spectrum of flux patches, but let us suppose as an example that the $4 \times 10^{16}$ Mx is made up of either two patches of flux of $2 \times 10^{16}$ Mx or 20 patches of $2 \times 10^{15}$ Mx. In the first case, each patch will have two opposite-polarity parts each having a size of 100 km, which take 100 s to cancel if the cancellation speed is 1 km s$^{-1}$. In the second case, each patch will release energy over a burst of duration only 30 s. If the field strength were instead 1 kG, the sizes and burst duration would be smaller by a factor of 3.

Suppose that each patch that is canceling consists of two parts of equal and opposite flux with a field strength $B_{100}$ in hundreds of Gauss, that the interface between the two polarities has dimension $\bar{D}_1$ Mm in the out-of-plane direction when

Table 1
Energy Release Rates for Different Size Cancellation Events

| No. Patches | Patch Flux | Size ($D$) | Timescale ($\tau$) | Energy per Event ($W$) | Overall Energy Release Rate |
|---|---|---|---|---|---|
| 20 | $2 \times 10^{15}$ Mx | 30 km | 30 s | $1.6 \times 10^{23}$ erg | $10^{22}$ erg s$^{-1}$ |
| 2 | $2 \times 10^{16}$ Mx | 100 km | 100 s | $1.6 \times 10^{24}$ erg | $10^{22}$ erg s$^{-1}$ [a] |

[a] Release ($W$) per cancellation event and overall energy release rate over the lifetime of a granule (300 s), for different configurations of flux patches that are consistent with the observed quiet Sun cancellation rate of Smitha et al. (2017), based on an assumed cancellation speed of 1 km s−1, field strength in the flux patches of 100 G, and current sheet length out of the symmetry plane of 100 km.

viewed from above, and that their in-plane horizontal extent is $D_1$ Mm. Then the energy that is released is

$$W = 2\frac{D^2 B^2 \bar{D}}{8\pi} = 0.8 \times 10^{27} D_1^2 \, B_{100}^2 \, \bar{D}_1 \text{ erg}. \quad (24)$$

Consider again the two example distributions of flux patches above (and see Table 1 for a summary). In the case of $2 \times 10^{16}$ Mx flux patches, if the field is $B = 100$ G, the size $D = 100$ km and the depth $\bar{D} = 200$ km, the energy release is $W = 1.6 \times 10^{24}$ erg, and the rate of energy release over the 300 s lifetime of the granule is $5 \times 10^{21}$ erg s$^{-1}$, or for two flux patches $10^{22}$ erg s$^{-1}$. These figures are for the whole granule, and they correspond to $5 \times 10^5$ and $10^6$ erg cm$^{-2}$ s$^{-1}$, respectively. Thus, comparing with Section 5.1, we see that the cancellation and precancellation phases should have comparable energy release rates. We note that a field strength 1 kG would instead give an energy release rate of $10^{23}$ erg s$^{-1}$. In the case of smaller $2 \times 10^{15}$ Mx flux patches of field strength $B = 100$ G, size $D = 30$ km, and depth $\bar{D} = 100$ km, the energy release is $W = 1.6 \times 10^{23}$ erg, but when summed over 20 patches, the rate of energy release is again $10^{22}$ erg s$^{-1}$ (or $10^{23}$ erg s$^{-1}$ for fields of 1 kG).

Now, in the quiet Sun, the energy required to heat the chromosphere and corona is $4 \times 10^6$ and $3 \times 10^5$ erg cm$^{-2}$ s$^{-1}$, respectively, or $4 \times 10^{22}$ and $3 \times 10^{21}$ erg s$^{-1}$, respectively, over the area ($10^{16}$ cm$^2$) of a granule. We conclude therefore that the observed rate of flux cancellation is likely to provide enough energy to heat the chromosphere and corona, and is similar in the two phases of precancellation and cancellation.

### 5.3. Phase 1 from the Flux Estimate

Finally, let us follow Longcope et al. (2001) and Priest et al. (2005) in estimating the energy that is dissipated in the precancellation phase when a current sheet forms in the corona above approaching flux patches. If the ambient field of strength $B_0$ contains a null point or separator at height $h_0^{3D}$, we suppose that as the field moves a distance $d$, say, a current sheet forms (Figure 3) and experiences reconnection of flux of amount

$$\Delta\psi = B_0 d \pi h_0^{3D}, \quad (25)$$

where $\pi h_0^{3D}$ is the extension of the current sheet out of the plane of the figure. Let us suppose that locally the field at the current sheet is $(B_0/h_0^{3D})(z^2 - \frac{1}{4}L^2)^{1/2}$, giving a field component along the sheet of $B_x = (B_0/h_0^{3D})(\frac{1}{4}L^2 - x^2)^{1/2}$ and a total current in the sheet of

$$I = \tfrac{1}{8} B_i \, L = I_0 \frac{L^2}{4(h_0^{3D})^2}, \quad (26)$$

where $B_i = B_0 L/(2h_0^{3D})$ is the inflow field to the current sheet and $I_0 = \tfrac{1}{4} h_0^{3D} B_0$.

Now, a second expression for the flux may be obtained by using the fact that the field on the inflow axis to the current sheet is $B_x(0, y) = (B_0/h_0^{3D})(\frac{1}{4}L^2 + y^2)^{1/2}$ and integrating the difference between this and the initial field $B_x(0, y) = B_0 y/h_0^{3D}$ near the null point to give, when $L \ll h_0^{3D}$,

$$\Delta\psi = \pi B_0 \int_0^{h_0^{3D}} \sqrt{\tfrac{1}{4}L^2 + y^2} - y \, dy$$
$$= \tfrac{1}{8}\pi B_0 L^2 \log(2h_0^{3D}/L). \quad (27)$$

Equating this to Equation (25) gives an expression for the sheet length of

$$\frac{L}{h_0^{3D}} = \left(\frac{16 d/h_0^{3D}}{-\log(d/h_0^{3D})}\right)^{1/2}, \quad (28)$$

where we assume $d < h_0^{3D}$ so as to make $-\log(d/h_0^{3D}) > 0$.

Finally, the energy release during reconnection of the flux $\Delta\psi$ may be written (Longcope et al. 2001) as

$$\Delta W = \tfrac{1}{2} I \, \Delta\psi, \quad (29)$$

which becomes, after substituting for $I$, $\Delta\psi$, and $L$ from Equations (26), (25), and (28), as well as $(h_0^{3D})^2 = F_0^{3D}/(\pi B_0)$:

$$\Delta W = \left(\frac{d/h_0^{3D}}{-2\log(d/h_0^{3D})}\right) \frac{d}{h_0^{3D}} h_0^{3D} B_0 F_0^{3D}. \quad (30)$$

Following the estimate in Section 5.2 of a typical flux cancellation of $4 \times 10^{16}$ Mx per granule, we put $F_0^{3D} = 2 \times 10^{16}$ Mx for one of the flux elements, as well as $B_0 = 10$ G and, say, $d = \tfrac{1}{2} h_0^{3D}$. Then the energy (30) liberated by reconnection becomes $10^{24}$ erg, which as we have seen above is sufficient to heat the corona above a granule during its lifetime.

## 6. Simulations of the Precancellation Phase

### 6.1. Simulation Setup

We wish to simulate the dynamics induced by the cancellation of a minority (or *parasitic*) polarity within a region containing one dominant polarity on the scale of a few granules. The flux from this dominant polarity may then form the flux within a coronal loop in the overlying atmosphere, or it may correspond to the open fields of coronal holes. Our simulation initial condition comprises a potential magnetic field in the volume that is generated by a magnetic bipole on the lower boundary surrounded by same-sign flux patches. The configuration is periodic in the $x$- and $y$-directions, so that the majority magnetic flux extends up to the top of the box. All of the magnetic flux from the parasitic polarity connects to surrounding dominant polarities. Between this low-lying "closed" flux and the flux that connects to the top of the domain is a separatrix surface associated with a magnetic null point. The separatrix forms a magnetic dome that arches down to the solar surface and separates the parasitic polarity from the surrounding unipolar region (Pontin & Priest 2022). This initial state is illustrated in Figure 5(a), and details of the magnetic field expression are provided in Appendix A.

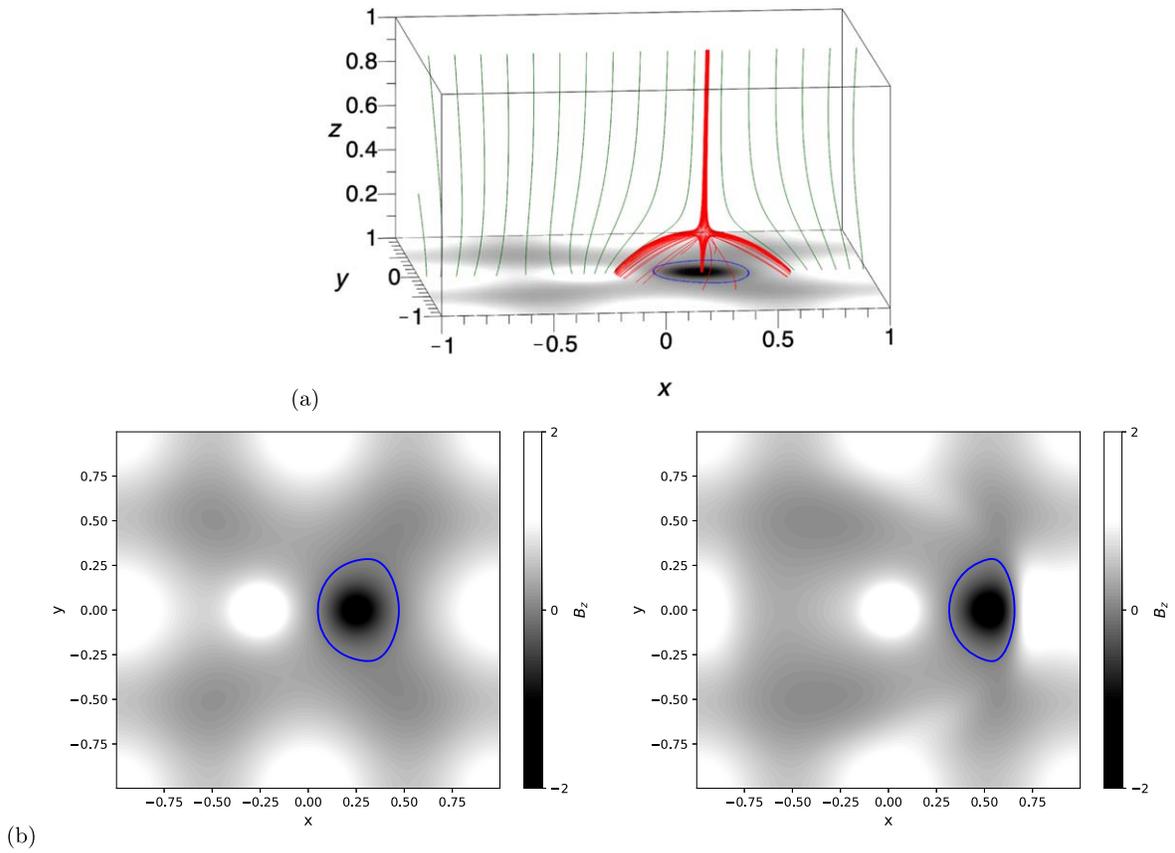

**Figure 5.** Initial conditions and boundary conditions for the simulation. (a) Magnetic field lines in the initial configuration, including some outlining the magnetic null points dome structure (red). (b) Distribution of $B_z$ at $z = 0$, with the polarity inversion line marked in blue, for $t = 0$ and $t = 7.15$ (left and right). Axes are labeled with the dimensionless units of the simulation, which can be converted to dimensional units as described in Section 6.1.

In order to induce flux cancellation, a tangential flow is imposed on the lower boundary that pushes the parasitic polarity toward the adjacent majority polarity flux patch, as shown in Figure 5(b). Explicitly, the prescribed flow is

$$v_x = v_0 \tanh^2(2t) \left[ 1 - \left( 1 - \cos\left(\frac{\pi x}{2}\right)^4 \cos\left(\frac{\pi y}{2}\right)^4 \right)^8 \right]. \quad (31)$$

This flow is prescribed at $z = 0$, with the domain spanning $x \in [-1, 1]$ Mm, $y \in [-1, 1]$ Mm, $z \in [0, 3.84]$ Mm. This domain is resolved with a grid of $512 \times 512 \times 768$ grid points, linearly spaced in $x$ and $y$ (with $\Delta x = \Delta y = 3.9$ km) and stretched in $z$ such that $\Delta z \approx 3.1$ km for $z \lesssim 1.5$ Mm, and gradually increases with $z$ above that level. On this grid, we use LaRe3D (Arber et al. 2001) to solve the single-fluid, resistive MHD equations including shock viscosity and thermal conduction. The code uses dimensionless units, which are adopted in many of the following figures, but for interpretation of the results we may set the units as follows: unit length $\equiv 1$ Mm, unit magnetic field strength $\equiv 0.01$ T (or 100 G), and unit density $\equiv 1.67 \times 10^{-7}$ kg m$^{-3}$. From these, all other dimensions follow.

In order to minimize magnetic field diffusion outside current sheets, we set the explicit resistivity ($\eta$) to zero throughout the domain except where the modulus of the current density ($|\mathbf{J}|$) exceeds a threshold value that is chosen to prevent the current sheet that forms from collapsing to the grid scale. Specifically, we set $\eta = 5 \times 10^{-4}$ where $|\mathbf{J}| > 25$ (in nondimensional code units), and zero otherwise.

In order to include thermal conduction at the correct level, the MHD equations are solved in dimensional form. For dimensionalization, we choose units of length to correspond to Mm, meaning that the domain represents the region above a (large) granule, and the parasitic polarity is on the scale of those observed by SUNRISE as reported by Smitha et al. (2017) and Chitta et al. (2017a). The magnetic field strength is chosen so that the field strength in the polarities has a maximum value of around 200 G. At $z = 2$ Mm, the average field strength is 25 G. This gives a flux for the parasitic polarity region (as well as the other flux patches) of $5 \times 10^{16}$ Mx, in the range of those observed by SUNRISE (Smitha et al. 2017). The plasma density and temperature are initially uniform, with values of $\rho = 1.67 \times 10^{-11}$ kg m$^{-3}$ and $2.3 \times 10^6$ K, respectively. As such, the lower boundary of our domain should be considered to be the base of the corona. A future study will consider the impact of atmospheric stratification on the results.

With the above choice of parameters, the sound speed in the domain is 200 km s$^{-1}$, while the Alfvén speed above the flux patches is $4 \times 10^3$ km s$^{-1}$. The driving speed is chosen to be below these two characteristic speeds, but is still faster than typically observed on the Sun, for computational expedience, we set $v_0 = 80$ km s$^{-1}$. This means that an interaction that would take minutes or hours to occur on the Sun takes only a few seconds, making simulation possible with the available computational resources.

### 6.2. Simulation Results

#### 6.2.1. Reconnection Onset, Flows, and Heating

The motion of the parasitic polarity region advects the spine footpoint of the coronal null. It is well established that such a motion drives the formation of a current sheet centered at the null point by a process of "null collapse" (e.g., Pontin & Craig 2005). The formation of the current sheet encourages *spine-fan reconnection* to take place (e.g., Pontin et al. 2007, 2013). This mode of reconnection involves a transport of magnetic flux and mass through the (fan) separatrix dome, and through/around the spine lines. In this configuration, specifically flux from the dominant polarity is transferred *into* the dome on its leading side (the side corresponding to its direction of motion), and an equivalent amount of flux is transferred *out of* the dome (i.e., is "opened up") on the trailing side. Because the flux of the parasitic polarity is approximately conserved, the flux transferred into the dome is equal and opposite to that transferred out. The direction of field-line motion in the symmetry plane is indicated by the arrows in Figure 6(a).

The flux is ejected from the reconnection site in two outflow jets (Figure 6(b)). One of these goes downward into the dome, gradually filling the leading lobe of the dome with hot plasma. The other is associated with the flux transfer out of the dome on its trailing side, with this jet being deflected upward along the "open" field lines into the upper part of the domain (Figure 6). In the symmetry plane ($y = 0$), the reconnection currents and flows resemble classical 2D reconnection configurations. Currents associated with slow-mode shocks bounding the outflow, as well as termination shocks (located at the ends of the reconnection outflow jets) and deflection currents, are clearly evident, and these are described in Appendix B.

The full 3D distribution of the flows and currents is shown in Figure 7. Figure 7(a) provides a simplified schematic, while frame (b) shows volume renderings of the data at two viewing angles, and frame (c) shows the flow in various cuts through the domain. Together, these illustrate that the current sheet is elongated in the "out-of-plane" ($y$) direction, being spread across the separatrix dome and extending all the way down to the photosphere—see also Appendix C. (Essentially the current flows upward along one flank of the dome, through the null—where it is most intense because the cross section of the sheet is at its shortest there—and then back down the other flank of the dome). Correspondingly, the upward outflow jet (green in Figures 7(a) and (b)) extends outward in the $y$-direction to form an upgoing ribbon-like structure of heated plasma.

The current sheet and reconnection process are quantified in the plots in Figure 8. These demonstrate that, once the current sheet forms, the reconnection is quasi-steady. In particular, the current sheet length in the symmetry plane ($y = 0$) stabilizes to a relatively steady value. As the flux patches approach progressively closer, the inflow magnetic field strength gradually increases, as does the inflow speed, meaning that the outflow speed and reconnection rate also gradually increase. Examining the maximum speed of the outflow jet (Appendix C) and comparing with the Alfvén speed in the inflow region in the symmetry plane, we find a good match between the two (within 10% or 20%). This match is exact for steady-state reconnection in the Sweet–Parker model, and is expected to be relatively close here because the reconnection occurs in a quasi-steady manner.

#### 6.2.2. Fluxes into the Upper Corona

In our model, the flux from the dominant magnetic polarity is intended to represent the flux at the footpoint of a coronal flux tube (that may be globally open or closed, extending up much higher into the corona than the upper boundary of our computational box). In order to quantify the effect of the reconnection process on the plasma at larger heights in the corona, we measure various fluxes at a height above the reconnection site. In this section, we choose the plane $z = 0.35$ to evaluate these fluxes and other quantities, because it is well above the reconnection site but also well away from the upper boundary (approaching which the upflow slows because $v = 0$ there).

As shown in Figure 7, the upward fluxes of mass and energy occur in a curved "ribbon"-type structure that contains the spine line of the null but extends away from it for some distance in the $y$-direction (i.e., the direction perpendicular to the driving flow). The Poynting flux (black line in Figure 8) grows approximately linearly in the early stage, before peaking and then settling to an approximately uniform value. The upward mass flux and thermal energy flux closely mirror each other: each has an initial rise, levels off, and then has additional growth (for $4.5 \lesssim t \lesssim 6$) that saturates by the end of the simulation. This second growth phase starting at $t \approx 4$ corresponds to the time when the plasma of the hot outflow jet first reaches this altitude.

It is worth noting that the simulations performed have a high degree of symmetry, which aids the analysis. To explore the effect of breaking this symmetry, further simulations were performed wherein the magnetic flux within the closed-field dome was twisted by a rotational boundary flow around the spine in a precursor phase, storing magnetic energy there. This changes the morphology of the evolution: the null point moves out of the $y = 0$ plane, while the current sheet that forms is no longer symmetric about $y = 0$, and therefore neither is the outflow jet. However, the quantitative energy fluxes into the upper part of the domain were within 5%–10% of those presented above, and thus we do not spend further time discussing these simulations.

#### 6.2.3. Energy Conversion

In our simulation, the parasitic polarity has a flux of $5 \times 10^{16}$ Mx and moves a distance of around 250 km at a speed of 80 km s$^{-1}$. Clearly, on the Sun, this would occur much more slowly, meaning that the process of energy release would be expected to take longer. In order to evaluate the total energy converted during the reconnection process, one should ideally calculate the Poynting flux into and out of the current sheet and associated shocks, and integrate over the time of the simulation. This is challenging to do in practice, due to the current sheet and shock geometry, and so instead we perform this analysis for a box of dimensions $0.08 \times 0.46 \times 0.06$ centered on the null and containing the central portion of the current sheet, and then multiply by a geometric factor to estimate the energy conversion in the whole sheet, giving $\sim 4 \times 10^{12}$ J or $4 \times 10^{19}$ erg. This is very much an underestimate of the total energy release expected from such an interaction on the Sun, because (i) the energy release will continue beyond the simulated interaction time, (ii) the energy conversion in the shocks is neglected, and (iii) at magnetic Reynolds numbers more appropriate for the corona, the reconnection onset would occur later, meaning greater Poynting flux injection through the photosphere, leading to more stored magnetic energy being available. For these reasons, it is not surprising that the above figure is two orders of magnitude below what would be predicted by Equation (24). Concerning the partition between the converted energy, we find that the reconnection process converts around three-fifths of the incoming Poynting flux into kinetic energy by doing work on the plasma (consistent with estimates from classical 2D reconnection models; Priest 2014)—see Figure 10. The other two-fifths remains for direct ohmic heating, although not all of this is captured by the explicit resistivity model employed, with some being lost to numerical dissipation.

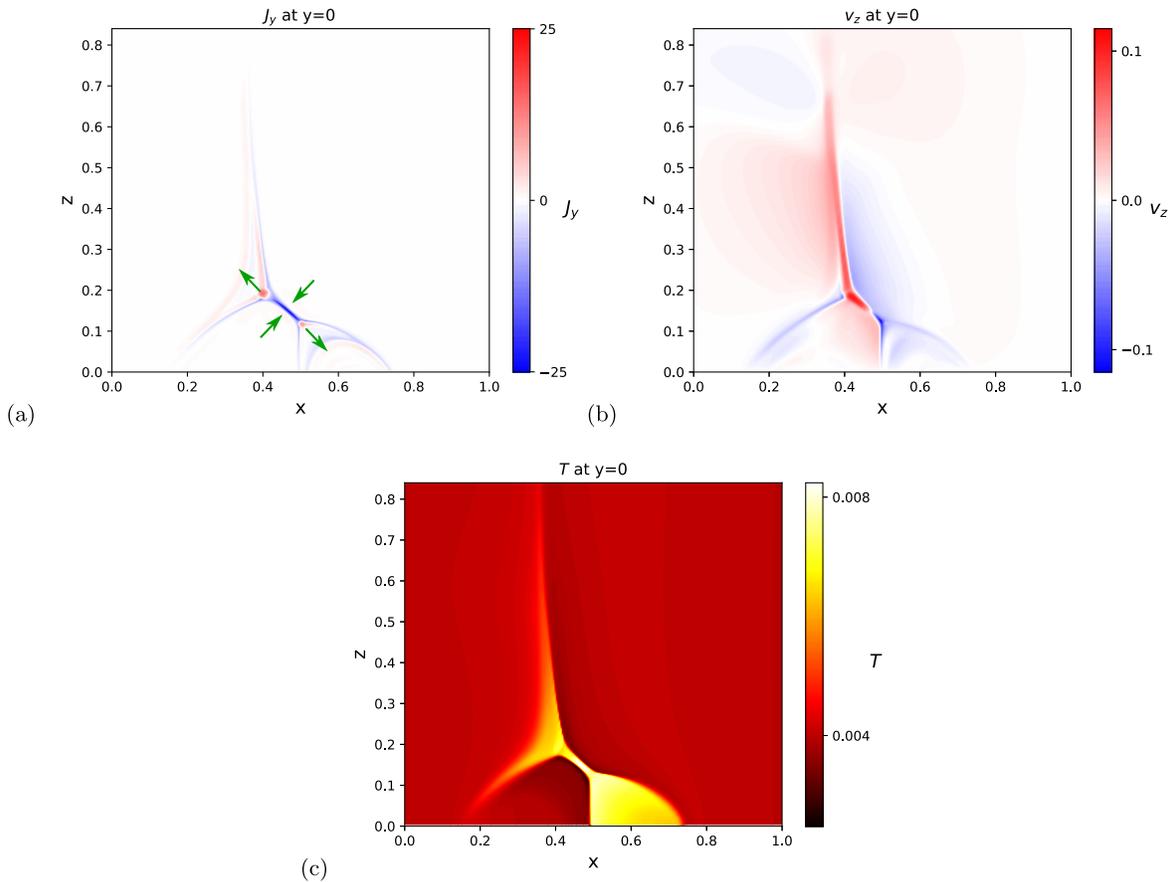

**Figure 6.** The current density (a), vertical velocity (b), and temperature (c) distribution in the $y=0$ plane at the end of the simulation. The green arrows in (a) indicate the direction of flux transfer at the reconnection site. Axes are labeled with the dimensionless units of the simulation, which can be converted to dimensional units as described in Section 6.1.

## 7. Discussion and Conclusions

### 7.1. Observable Signatures

During a flux cancellation event such as those modeled above, we have seen that both local plasma heating and the formation of a hot upward jet occur (on the Sun, it is likely some nonthermal acceleration will also occur, but this is beyond the scope of our models). In the case considered here, the energy release is mediated by a current sheet that forms about a 3D null, but as illustrated in Figure 1, the current sheet may also form at a separator, e.g., if the overlying flux is locally horizontal (Priest & Syntelis 2021). Such events will lead to a local increase in emission intensity, which may exhibit a "spire"-type geometry due to the upflowing jet—see Figures 2, 6, and 7. It is noteworthy that the observation of such a hot jet could be highly dependent on the line of sight, given its highly anisotropic cross section (see Figure 7). The timescale of the emission enhancement will be on the order of the flux cancellation time. Based on SUNRISE observations, Anusha et al. (2017) found that small-scale magnetic patches have a median lifetime of 66 s. This lifetime estimate is obviously limited by the cadence of magnetograms. If we assume that it also reflects the timescale of cancellation, then these timescales are comparable to the timescales of EUV emission variations in campfires in the corona (Berghmans et al. 2021). Such short timescales are also found in the jets displayed in Figure 2.

### 7.2. Atmospheric Heating

The analytical models allow us to estimate the energy release rate for cancellation events of different sizes and different topologies. The present work complements the previous theory of flux cancellation in a horizontal field by focusing on the technically trickier aspect of its nature in a vertical field. The energy release rate is found to be sufficient to heat the chromosphere and corona, on the basis of the latest estimates of flux cancellation rates. The MHD simulations confirm the conversion of energy in reconnecting current sheets, in a geometry representing a small-scale bipole being advected toward an intergranular lane.

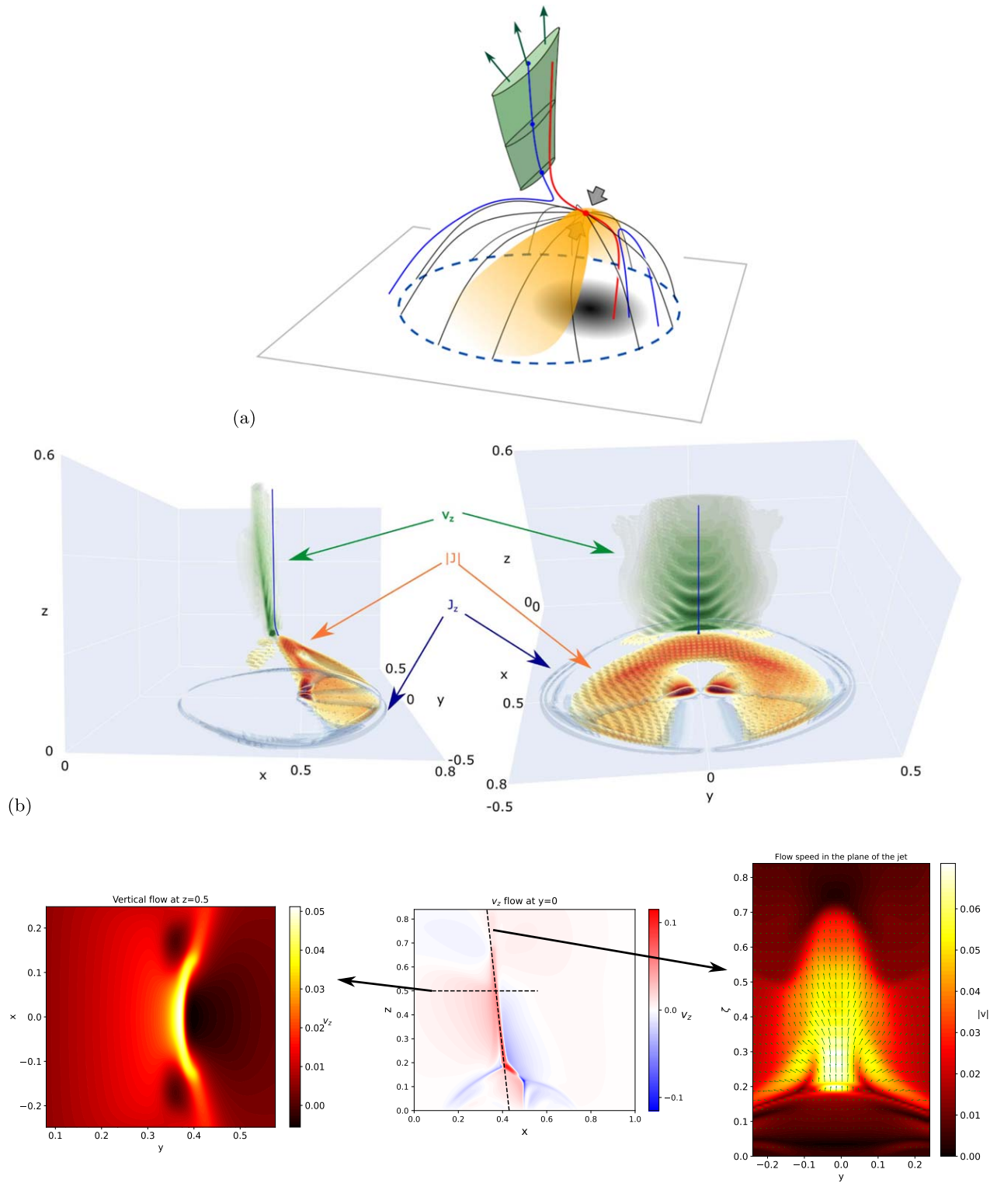

**Figure 7.** (a) Schematic three-dimensional view of the upflowing mass and energy at the end of the simulation. Orange shows the location of the current, and green the upward jet. The field lines of the dome (black) close down along the dashed blue curve. (b) From two different viewing angles, at the final time of the simulation, volume rendering of the current density (yellow–red). Also shown is the flow component $v_z$ (green) restricted to the volume above the null. The blue shading on the base shows $J_z$ there, outlining the footprint of the dome, and the blue curve marks the "outer spine." (c) Cuts through the upgoing jet illustrating how the flow varies out of the symmetry plane. Axes are labeled with the dimensionless units of the simulation, which can be converted to dimensional units as described in Section 6.1.

Our basic thesis is that the small-scale reconnection events are locally very similar in a globally closed or open magnetic field, with the magnetic energy release showing up as heating, kinetic energy, conductive flux, fast particles, and due to the time-dependent nature of the energy conversion, as waves. However, their global impact due to different energy pathways is quite different. In the atmosphere of a magnetically closed field above, say, a supergranule cell, all of the different energy components are subsequently degraded into heat and contribute to atmospheric heating, with the exception of fast-mode waves that escape the closed region. However, for a magnetically open supergranule cell, much of the energy is available for accelerating the solar wind and creating the small-scale structuring observed in the solar wind, either directly as jetlets or indirectly via waves or fast particles. This implies that less energy is available for heating a coronal hole.

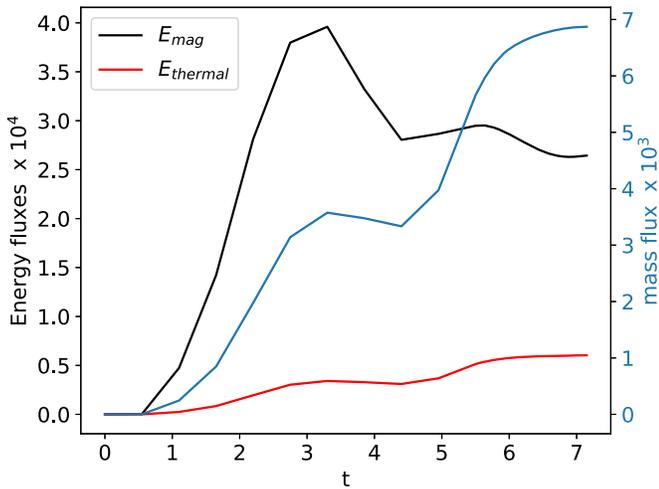

**Figure 8.** Flux of magnetic energy (black), thermal energy (red), and mass (blue) in the jet as a function of time, at height $z = 0.35$. Axes are labeled with the dimensionless units of the simulation, which can be converted to dimensional units as described in Section 6.1.

One shortcoming of our simulation approach is that the actual cancellation of the two flux patches has not yet been modeled. This would require incorporating a downflow through the lower boundary and a corresponding reduction in the flux of the parasitic polarity, as well as a modeling of the lower atmosphere. We have therefore terminated our simulations when the separation of the opposite-polarity flux patches becomes too small, and can only assess the fluxes into the upper domain up until that time. It is important to note that there are many caveats to the energy estimates in the simulations, given the simplified nature of our modeling. One example of this is the absence of atmospheric stratification, such that the reconnection site associated with the flux cancellation is already in the corona. The effect of different heights for the reconnecting current sheet will be explored in a follow-up work. Another simplification is the absence of radiative losses in the energy equation. The thermal structure of the loops in the outflow region on the leading side of the dome will in reality be governed by evaporation of the photospheric material, its motion upward into the corona, and radiation losses of the evaporated dense material (as in, e.g., the jet simulations of Lionello et al. 2016).

### 7.3. Conclusions

We conclude that through two phases of atmospheric energy release—precancellation and cancellation—the cancellation of photospheric magnetic flux fragments may provide a significant contribution to coronal heating and solar wind generation. The analytical models predict that a comparable energy release rate should occur in each of these two phases, although it is not yet known whether or not the release by cancellation in the second phase can escape the photosphere, a result that should be tested against observations and advanced simulation work. Based on the newly measured rates of flux cancellation (Smitha et al. 2017), the energy released by flux cancellation is estimated from our models to be of the order $10^6 \, \mathrm{erg\, cm^{-2}\, s^{-1}}$, which compares favorably with the energy required to heat the chromosphere and corona. Thus, we conclude that reconnection driven by flux cancellation may play a significant role in heating the solar atmosphere and (when it is interchange reconnection in a globally open field region) generating the solar wind, an idea supported by new observations, e.g., of transition-region upflows (Tripathi et al. 2021), jetlets (Raouafi et al. 2023) and active-region loops (Chitta et al. 2018, 2020).

In future, there are many observational and theoretical aspects to explore. Observationally, it will be important to study the effect on the overlying atmosphere of the cancellation phase of the process when the canceling features are in contact and decreasing in flux. In addition, more details of cancellation at the smallest scales would be invaluable, including the effect of the neighboring magnetic topology, the flux sizes, the field strength, and the speed of approach on the energy release. On the theoretical side, stratification needs to be added, and more details need to be studied on the different energy pathways, on the cancellation phase, and on the observational implications.

### Acknowledgments

E.R.P. is thankful to Clare Parnell for insightful discussions. D.I.P. gratefully acknowledges support through an Australian Research Council Discovery Project (DP210100709) and helpful discussions with J. Reid, L. Tarr, J. Leake, and L. Daldorff. L.P.C. gratefully acknowledges funding by the European Union (ERC, ORIGIN, 101039844). Views and opinions expressed are, however, those of the author(s) only and do not necessarily reflect those of the European Union or the European Research Council. Neither the European Union nor the granting authority can be held responsible for them. V.S.T.'s contribution was supported by NASA grants 80NSSC20K1317, 80NSSC22K1021 and 80NSSC20K0192, and NSF grant ICER1854790. Computations were run on the Australian National Computational Infrastructure's Gadi machine through an award from Astronomy Australia Ltd.'s Astronomy Supercomputer Time Allocation Committee. Solar Orbiter is a space mission of international collaboration between ESA and NASA, operated by ESA. We are grateful to the ESA SOC and MOC teams for their support. The EUI instrument was built by CSL, IAS, MPS, MSSL/UCL, PMOD/WRC, ROB, and LCF/IO with funding from the Belgian Federal Science Policy Office (BELSPO/PRODEX PEA 4000112292 and 4000134088); the Centre National d'Etudes Spatiales (CNES); the UK Space Agency (UKSA); the Bundesministerium für Wirtschaft und Energie (BMWi) through the Deutsches Zentrum für Luft- und Raumfahrt (DLR); and the Swiss Space Office (SSO). This research has made use of NASA's Astrophysics Data System Bibliographic Services.

## Appendix A
## Initial Magnetic Field for the Simulations

The initial magnetic field takes the form of an array of submerged dipoles, superimposed upon a uniform background field. The uniform background field is $\boldsymbol{B} = -0.8\hat{\boldsymbol{z}}$, and the magnetic field of each dipole is defined by

$$\boldsymbol{B}_d = B_0 \nabla \times \left( \frac{\hat{\boldsymbol{z}} \times \hat{\boldsymbol{r}}_0}{\boldsymbol{r}_0 \cdot \boldsymbol{r}_0} \right),$$
$$\boldsymbol{r}_0 = (x - x_0)\hat{\boldsymbol{x}} + (y - y_0)\hat{\boldsymbol{y}} + (z - z_0)\hat{\boldsymbol{z}}, \qquad (A1)$$

where the hat notation denotes a unit vector. Eight dipoles are placed around the edges of the domain $x, y = \pm 1$ with a strength of $B_0 = +1$ at a depth $z_0 = -1$. An additional four dipoles placed at $(1/4, 1/4, -1)$, $(-1/4, 1/4, -1)$, $(1/4, -1/4, -1)$, $(-1/4, -1/4, -1)$, and with $B_0 = 0.6$ create a weak positive background polarity at the photosphere.

The pair of opposite polarities close to the origin are created using a pair of submerged monopoles:

$$\boldsymbol{B}_d = B_1 \left( \frac{\boldsymbol{r}_0}{(\boldsymbol{r}_0 \cdot \boldsymbol{r}_0)^{3/2}} \right). \qquad (A2)$$

One of these has $B_1 = 0.05$ and is located at $(-1/4, 0, -1/6)$, and the other has $B_0 = -0.05$ and is placed at $(1/4, 0, -1/6)$. We note that this magnetic field (as well as $B_0$ and $B_1$) is dimensionless—as described in Section 6.1, when dimensionalized it has units of 100 G. In order to produce an initial condition compatible with periodic boundary conditions in $x$ and $y$ (for computational expedience), we replicate this boundary distribution over a $9 \times 9$ grid extending over $x, y \in [-9, 9]$, but simulate only the domain $x, y \in [-1, 1]$. The resulting boundary distribution of magnetic field strength and representative field lines in the initial condition are shown in Figure 5.

## Appendix B
## Current Layer and Shock Geometry

As shown in Figure 6, the current layer geometry in the symmetry plane is reminiscent of a classical X-point reconnection configuration (the 3D extension being shown in Figure 7). If we focus for simplicity on this symmetry plane, we see a number of features that have been noted in the 2D reconnection literature. These include prominent reverse-current spikes at each outflow end of the current layer, and four "fronts" of high current density extending from the four corners of the current layer (Figure 9). It is worth noting that the presence of these features is highly dependent on the resistivity model employed. In these simulations, the resistivity is localized within the region of high current density (see Section 6.1)—in test runs with uniform resistivity, these features were not present (or at least much less obviously so). The four current fronts emanating from the current layer are often described as "separatrix currents," but we can see that this is a misnomer by examining Figure 9, where it is clear that the fronts and the field lines are not exactly aligned, such that the current fronts are not coincident with magnetic flux surfaces, in particular the separatrix (fan) or spine.

In Petschek's model of reconnection, there are four standing slow-mode shocks extending from the corners of the current layer. Examining the variation of the physical variables across the current fronts in our simulations, it is clear that they are the 3D analogs of these structures. In particular, both the density and temperature increase from the upstream to downstream side of the current front, while the modulus of $\boldsymbol{B}$ decreases, indicating that the field vector is rotated away from the normal as the front—or shock—is crossed. We note that the jet and current are downstream of the separatrix, as expected for a slow-mode shock.

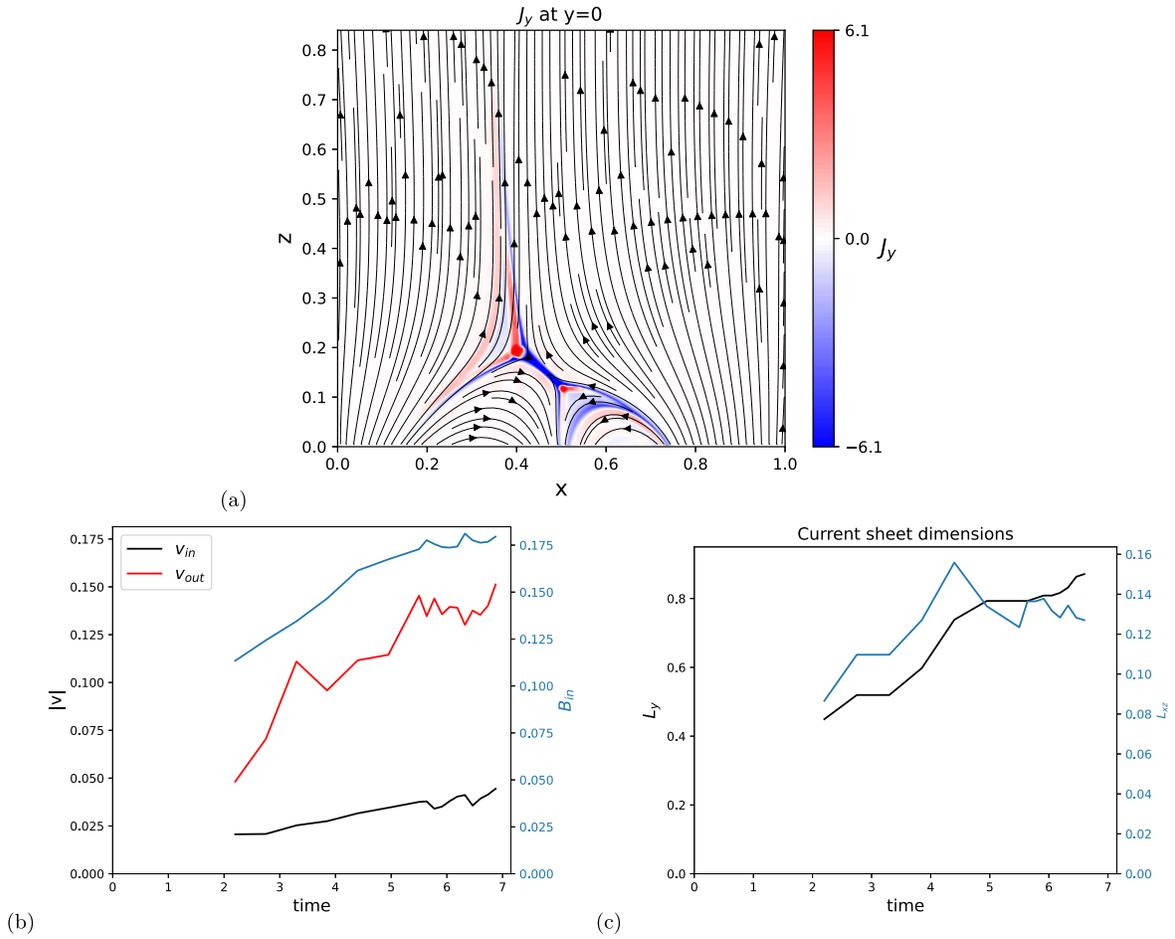

**Figure 9.** (a) Current density in the $y = 0$ plane (saturated to $\pm 6$), overlaid with streamlines of the magnetic field. The current fronts (∼shocks) are not coincident with magnetic field lines/flux surfaces, in particular the separatrix (fan) or spine. This is particularly evident in the vicinity of the outer spine. (b) Components of the flow velocity $\boldsymbol{v}$ and magnetic field, $\boldsymbol{B}$, either parallel or perpendicular to the sheet axis, measured in the $y = 0$ plane as the difference between the values on either side of the current sheet. (c) Dimensions of the current sheet as a function of time. $L_{xz}$ (blue) is the length of the sheet in the $y = 0$ plane. $L_y$ is the linear distance between current sheet "ends" in the $y$-direction, where the end is defined by the point at which $|J|$ falls to 50% of its maximum value. The asymptotic value of ∼0.8 is approximately the width of the dome, indicating that the current sheet has spread all across the dome down to the lower boundary. Axes are labeled with the dimensionless units of the simulation, which can be converted to dimensional units as described in Section 6.1.

## Appendix C
## Current Layer Dimensions and Inflow Properties

Figure 9(c) shows the current sheet dimensions defined as follows: in the $xz$-plane, we identify the outflow directions and find the distance in each direction by which $|\boldsymbol{J}|$ drops to 10% of the peak value. For the extension in the $y$-direction, i.e., out of plane, this is more difficult because of the current layer curvature, etc. Shown in Figure 9(c) is the linear distance along $y$ between the current sheet end points defined either as the points where the current density falls to 50% of its maximum. This occurs all the way down at the photosphere at later times, as the current layer grows until the current reaches the chosen value at the photosphere.

For the partition of energy conversion in the current layer, see Figure 10. Consistent with 2D models approximately three-fifths of the incoming magnetic energy (calculated via the Poynting flux) is converted to kinetic energy, calculated with the work term $(\boldsymbol{J} \times \boldsymbol{B}) \cdot \boldsymbol{v}$. The rest of the energy should be converted to heat via ohmic dissipation. The fact that this does not match (panel (b)) indicates that the current layer is not perfectly resolved (probably where $J$ is below the threshold for $\eta$ to switch on), so the heating is underestimated (by a factor of ∼2 for $t \in [5, 7]$).

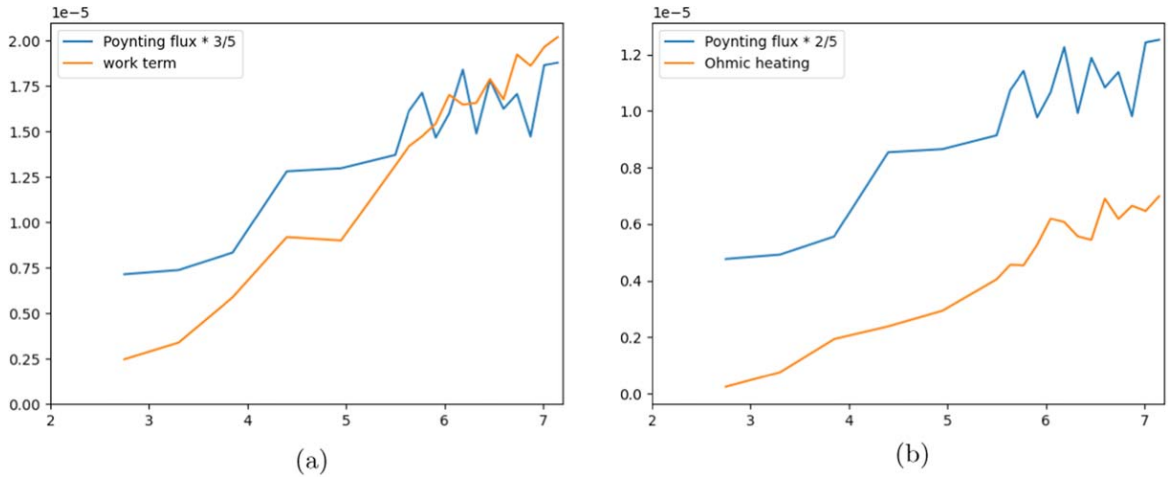

**Figure 10.** Poynting flux into a box of size $0.08 \times 0.46 \times 0.06$ (dimensionless units) centered on the null, compared to (a) the work done by the magnetic field on the plasma and (b) the ohmic dissipation, in the same box. Axes are labeled with the dimensionless units of the simulation, which can be converted to dimensional units as described in Section 6.1.

## Appendix D
## Choice of Flux Conservation Condition

Consider the initial 2D situation with the negative source $S$ at the origin and a null point above it, as shown in Figure 11(a), where the separatrix reaches the surface in the points $P_1$ and $P_2$, and there are four field lines that approach the source from each side below the separatrix. Then suppose the source moves to the right and a current sheet forms with no reconnection. The flux above the separatrix will be conserved so that the fluxes across $\mathcal{C}2$ and $\mathcal{C}4$ in Figure 4 will be conserved. But what about the fluxes below the separatrix?

At first sight, the natural possibility is to insist that the model is strictly two-dimensional, so that, as the source $S$ moves, the base fluxes between $P_1$ and $S$ and between $S$ and $P_2$ are conserved, which in turn implies that the field-line footpoints between $P_1$ and $S$ will move further apart and the field strength will decrease, while those between $S$ and $P_2$ will become closer together and the field strength will increase (Figure 11(b)). This would imply that the fluxes below the current sheet along $\mathcal{C}1$ and $\mathcal{C}3$ are conserved.

However, a second possibility, which is what we adopt here, is to remember that, in three dimensions as viewed from above, the situation is as seen in Figures 11(d) and (e), so that the field-line connections of the source to footpoints in the surrounding region are all conserved. This implies that initially there are the same number of field lines stretching out to the left of $S$ and to the right of $S$, whereas after the source has moved to the right, there are more field lines to the left than to the right. At the same time, the normal magnetic field component over most of the photosphere has remained constant. Thus, the best way of reproducing this effect in a 2D model is to assume that, outside the source $S$, the normal magnetic field remains constant, as in Figure 11(c). This implies that there are now more magnetic field lines to the left of $S$ than to the right. In other words, as the main source $S$ moves, we assume the little sources of the uniform background field that are ahead of the source simply move around the main source and so lie behind it, as they would appear to do so when viewed from the side in three dimensions.

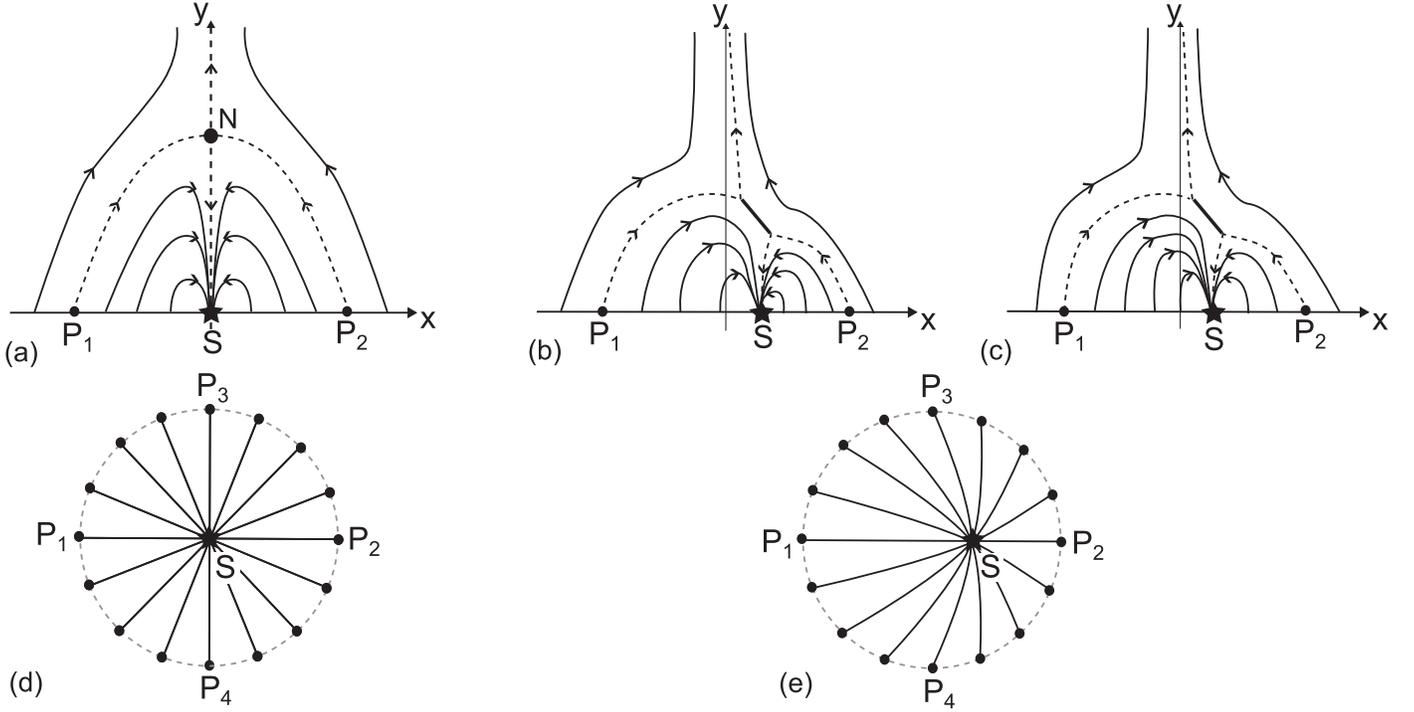

**Figure 11.** The analytical model, showing (a) the initial 2D configuration, (b) the configuration after the source $S$ has moved to the right with no reconnection when the magnetic flux at the base to the left and right of the source have remained constant, (c) the corresponding configuration when the values of the normal field component at the base to the left and right of the source have remained constant, (d) the initial 3D configuration viewed from above, and (e) the result after the source has moved.

## Appendix E
## Evaluation of the Complex Function $P(z)$

We have from Equations (8) and (E3) that

$$B_y + iB_x = \left[(z - z_1)(z - \bar{z}_1)(z - z_2)(z - \bar{z}_2)\right]^{1/2} P(z), \quad (E1)$$

where

$$P(z) = -\frac{i}{\pi} \int_{-\infty}^{\infty} \frac{B_y(\xi, 0)\, d\xi}{(\xi - z)[(\xi - z_1)(\xi - \bar{z}_1)(\xi - z_2)(\xi - \bar{z}_2)]^{1/2}}, \quad (E2)$$

where $B_y(\xi, 0) = B_0(1 - \pi h_0 \delta(\xi - d))$.

After substituting for $B_y(\xi, 0)$ in the expression for $P(z)$, it becomes

$$B_y + iB_x = -\frac{iB_0}{\pi}\left[(z - z_1)(z - \bar{z}_1)(z - z_2)(z - \bar{z}_2)\right]^{1/2}(I_1 + I_4), \quad (E3)$$

where

$$I_1 = \int_{-\infty}^{\infty} \frac{d\xi}{(\xi - z)\left[(\xi - z_1)(\xi - \bar{z}_1)(\xi - z_2)(\xi - \bar{z}_2)\right]^{1/2}} \quad (E4)$$

and

$$I_4 = -\frac{\pi h_0}{\left[(d - z_1)(d - \bar{z}_1)(d - z_2)(d - \bar{z}_2)\right]^{1/2}}. \quad (E5)$$

### E.1. The End Points of a Short Current Sheet

In order to evaluate $I_1$, we assume the current sheet is short by expanding its endpoints $z_1$ and $z_2$ in powers of the distance $d$ moved by the sources, where $d \ll h_0$ and $L \ll h_0$. So we write

$$z_1 = z_0 + 2\sqrt{dh_0}\, u, \quad \text{and} \quad z_2 = z_0 - 2\sqrt{dh_0}\, u, \quad (E6)$$

where the position $z_0$ of the center of the sheet is

$$z_0 = ih_0 + dv, \quad (E7)$$

and

$$L = 4\sqrt{dh_0}\, u. \quad (E8)$$

Here, the complex unknowns $v$ and $u$ are to be determined from asymptotic and flux conditions.

First, the integrand in the expression for $I_1$ may be expanded in powers of $d$ up to the linear term, to give

$$\frac{1}{(\xi - z)(\xi^2 + h_0^2)} + \frac{(a_3\xi^3 + a_2\xi^2 h_0 + a_1\xi h_0^2 + a_0 h_0^3)d}{(\xi - z)\left(\xi^2 + h_0^2\right)^3}, \quad (E9)$$

where

$$a_3 = v + \bar{v}, \quad a_2 = iv - i\bar{v} + 2u^2 + 2\bar{u}^2,$$
$$a_1 = v + \bar{v} + 4iu^2 - 4i\bar{u}^2$$
$$a_0 = iv - i\bar{v} - 2u^2 - 2\bar{u}^2. \quad (E10)$$

Then performing the integral $I_1$ gives

$$I_1 = -\frac{\pi}{h_0(z + ih_0)} + I_1^{(1)}, \quad (E11)$$

where

$$I_1^{(1)} = \frac{d}{8(iz-h_0)^3 h_0^3}[-ia_3(3iz-h_0)h_0^2 \\ - a_2(9ih_0z - 8h_0^2 + 3z^2)h_0 \\ - ia_1(iz - 3h_0)h_0^2 \\ + i(iz - 3h_0)z(i\bar{v} - iv - 2\bar{u}^2 - 2u^2)]. \quad (E12)$$

The asymptotic behavior of this integral as $z$ approaches infinity is

$$I_1 \sim \left\{ \frac{3\bar{v} + v + i\bar{u}^2 + iu^2}{2z^3} \right. \\ \left. + \frac{i\bar{v} + iv + \bar{u}^2 - u^2}{2h_0 z^2} + \frac{\bar{v} - v - i\bar{u}^2 - iu^2}{2h_0^2 z} \right\}. \quad (E13)$$

This may be substituted into the Equation (E1) for $B_y + iB_x$, with $P(z)$ given by Equation (E3) and compared with the required asymptotic behavior, namely

$$B_y + iB_x \sim B_0\left(1 - \frac{ih_0}{z}\right), \quad (E14)$$

to give the following expressions for the unknowns $u$ and $v$ in terms of a third real unknown $w$:

$$v = 1 - w^2, \quad u = \frac{1-i}{\sqrt{2}}w. \quad (E15)$$

The corresponding expressions for the end points $z_1$ and $z_2$ of the current sheet follow from Equations (E6) and (E7).

### E.2. Flux Function

Our initial magnetic field is from Equation (7)

$$\mathcal{B}_0(z) = B_0\left(1 - \frac{ih_0}{z}\right), \quad (E16)$$

where the initial null point is located at $z_0 = ih_0$. Also, after expanding the integrals $I_1$ and $I_4$ in powers of $d$ to order $d$, the expression for the magnetic field after the source has moved a distance $d$ may be written in the form

$$\mathcal{B}(z) \equiv B_y + iB_x = -\frac{B_0\sqrt{[z-\tilde{z}_0]^2 + 4iw^2d}}{ih_0 + d} \\ \times \left(\frac{ih_0 + (1-w^2)d}{z-d} + \frac{w^2d}{z+ih_0}\right), \quad (E17)$$

where the center of the current sheet is at $\tilde{z}_0 = ih_0 + (1-w^2)d$ and its endpoints are

$$z_1 = \tilde{z}_0 + \sqrt{2dh_0}(1-1)w \text{ and} \\ z_2 = \tilde{z}_0 - \sqrt{2dh_0}(1-1)w. \quad (E18)$$

Our approach for finding the final piece of the jigsaw puzzle, namely an expression for the unknown $w$, is to insist that, if there is no reconnection as the current sheet builds up, then the values of the flux function $A$ are the same at the center $\tilde{z}_0$ of the current sheet and at the initial null point $z_0$, so that they lie on the same field line. The initial complex flux function is given by

$$\mathcal{A}_0(z) = \int_z^{z_0} \mathcal{B}_0 dz \quad (E19)$$

if we set its value at the initial null point $z_0$ to zero. On the other hand, the complex flux function with the current sheet is given by $\mathcal{A}(z) = \int_z^{z_0} \mathcal{B}(z)\,dz$, and so its value at the center of the current sheet may be approximated when $d \ll h_0$ using the trapezium rule obtained from Taylor's theorem as

$$\mathcal{A}(\tilde{z}_0) = \int_{\tilde{z}_0}^{z_0} \mathcal{B}(z)\,dz \approx \tfrac{1}{2}[\mathcal{B}_0(z_0) + \mathcal{B}(\tilde{z}_0)](\tilde{z}_0 - z_0), \quad (E20)$$

where $\tilde{z}_0 - z_0 = (1-w^2)d$. Indeed, after calculating $\mathcal{B}_0(z_0) + \mathcal{B}(\tilde{z}_0)$ to lowest order in $d$ and substituting into Equation (E20), we find the difference between the real flux functions to be

$$\text{Re}(\mathcal{A}(\tilde{z}_0) - \mathcal{A}_0(z_0)) = B_0 h_0 \sqrt{2}\,w(1-w^2)\frac{d^{3/2}}{h_0^{3/2}}. \quad (E21)$$

Thus, the values of the flux function at the initial null point and the center of the current sheet are the same, provided

$$w^2 = 1. \quad (E22)$$

In turn this implies that the end points of the sheet are located at $z_1 = ih_0 + \sqrt{2dh_0}\,(1-i)$ and $z_2 = ih_0 - \sqrt{2dh_0}\,(1-i)$, so that, to order $d$, as the source moves a small distance $d$, the sheet is centered at $ih_0$, is inclined at $\pi/4$, and has a length of

$$L = 2|z_1 - z_2| = 4\sqrt{dh_0}. \quad (E23)$$

### ORCID iDs

D. I. Pontin ● https://orcid.org/0000-0002-1089-9270
E. R. Priest ● https://orcid.org/0000-0003-3621-6690
L. P. Chitta ● https://orcid.org/0000-0002-9270-6785
V. S. Titov ● https://orcid.org/0000-0001-7053-4081